\input harvmac   
\noblackbox   


\def\p{\partial}

\def\CM{{\cal M}}    
\def\CN{{\cal N}}

\def\CF{{\cal F}}

\def\CV{{\cal V}}

\def\CS{{\cal S}}    
\def\CA{{\cal A}}

\def\CV{{\cal V }}    
    
\def\CS{{\cal S }}

\def\Tr{{\rm Tr}}

\def\zb {\bar{z}}    
\def\mb {\bar{m}}
 \def\nb {\bar{n}}


\font\manual=manfnt     
\def\dbend{\lower3.5pt\hbox{\manual\char127}}

\def\IZ{\relax\ifmmode\mathchoice    
{\hbox{\cmss Z\kern-.4em Z}}{\hbox{\cmss Z\kern-.4em Z}}    
{\lower.9pt\hbox{\cmsss Z\kern-.4em Z}} {\lower1.2pt\hbox{\cmsss    
Z\kern-.4em Z}}\else{\cmss Z\kern-.4em Z}\fi}    
\def\half {{1\over 2}}

\def\p{\partial}    
\def\pb{\bar{\partial}}    
\def\bar{\overline}    
\def\CS{{\cal S}}    
\def\CN{{\cal N}}    
    
\def\rt2{\sqrt{2}}    
\def\irt2{{1\over\sqrt{2}}}

\def\t{\tilde}    
\def\ndt{\noindent}    
\def\s{\sigma}    
\def\b{\beta}    
\def\a{\alpha}

\font\cmss=cmss10    
\font\cmsss=cmss10 at 7pt    
\def\IL{\relax{\rm I\kern-.18em L}}    
\def\IH{\relax{\rm I\kern-.18em H}}    
\def\IR{\relax{\rm I\kern-.18em R}}    
\def\inbar{\vrule height1.5ex width.4pt depth0pt}    
\def\IC{\relax\hbox{$\inbar\kern-.3em{\rm C}$}}    
\def\rlx{\relax\leavevmode}    
\def\ZZ{\rlx\leavevmode\ifmmode\mathchoice{\hbox{\cmss Z\kern-.4em Z}}    
 {\hbox{\cmss Z\kern-.4em Z}}{\lower.9pt\hbox{\cmsss Z\kern-.36em Z}}    
 {\lower1.2pt\hbox{\cmsss Z\kern-.36em Z}}\else{\cmss Z\kern-.4em    
 Z}\fi}     
\def\IZ{\relax\ifmmode\mathchoice    
{\hbox{\cmss Z\kern-.4em Z}}{\hbox{\cmss Z\kern-.4em Z}}    
{\lower.9pt\hbox{\cmsss Z\kern-.4em Z}}    
{\lower1.2pt\hbox{\cmsss Z\kern-.4em Z}}\else{\cmss Z\kern-.4em    
Z}\fi}    
    

\def\Tr{{\rm Tr}}

\def\zb {\bar{z}}

\def\mb {\bar{m}} 
\def\mbar {\bar{m}}

\font\manual=manfnt     
\def\dbend{\lower3.5pt\hbox{\manual\char127}}

\def\IZ{\relax\ifmmode\mathchoice    
{\hbox{\cmss Z\kern-.4em Z}}{\hbox{\cmss Z\kern-.4em Z}}    
{\lower.9pt\hbox{\cmsss Z\kern-.4em Z}} {\lower1.2pt\hbox{\cmsss    
Z\kern-.4em Z}}\else{\cmss Z\kern-.4em Z}\fi}    
\def\half {{1\over 2}}

\def\pb{\bar{\partial}}    
\def\bar{\overline}    

\def\rt2{\sqrt{2}}    
\def\irt2{{1\over\sqrt{2}}}

\def\t{\tilde}    
\def\T{\widetilde}
\def\ndt{\noindent}    
\def\s{\sigma}    
\def\bB {{\bf B}}

\def\bra#1{{\langle}#1|}
\def\ket#1{|#1\rangle}

\let\includefigures=\iftrue
\let\useblackboard=\iftrue
\newfam\black

\includefigures
\message{If you do not have epsf.tex (to include figures),}
\message{change the option at the top of the tex file.}
\input epsf
\def\figin{\epsfcheck\figin}\def\figins{\epsfcheck\figins}
\def\epsfcheck{\ifx\epsfbox\UnDeFiNeD
\message{(NO epsf.tex, FIGURES WILL BE IGNORED)}
\gdef\figin##1{\vskip2in}\gdef\figins##1{\hskip.5in}
\else\message{(FIGURES WILL BE INCLUDED)}%
\gdef\figin##1{##1}\gdef\figins##1{##1}\fi}
\def\DefWarn#1{}
\def\figinsert{\goodbreak\midinsert}
\def\ifig#1#2#3{\DefWarn#1\xdef#1{fig.~\the\figno}
\writedef{#1\leftbracket fig.\noexpand~\the\figno}%
\figinsert\figin{\centerline{#3}}\medskip\centerline{\vbox{
\baselineskip12pt\advance\hsize by -1truein
\noindent\footnotefont{\bf Fig.~\the\figno:} #2}}
\bigskip\endinsert\global\advance\figno by1}
\else
\def\ifig#1#2#3{\xdef#1{fig.~\the\figno}
\writedef{#1\leftbracket fig.\noexpand~\the\figno}%
\global\advance\figno by1}
\fi

\def\doublefig#1#2#3#4{\DefWarn#1\xdef#1{fig.~\the\figno}
\writedef{#1\leftbracket fig.\noexpand~\the\figno}%
\figinsert\figin{\centerline{#3\hskip1.0cm#4}}\medskip\centerline{\vbox{
\baselineskip12pt\advance\hsize by -1truein
\noindent\footnotefont{\bf Fig.~\the\figno:} #2}}
\bigskip\endinsert\global\advance\figno by1}


\lref\mcgreevy{
  J.~McGreevy and H.~Verlinde,
  ``Strings from tachyons: The c = 1 matrix reloaded,''
  JHEP {\bf 0312}, 054 (2003)
  [arXiv:hep-th/0304224] \semi
  I.~R.~Klebanov, J.~Maldacena and N.~Seiberg,
  ``D-brane decay in two-dimensional string theory,''
  JHEP {\bf 0307}, 045 (2003)
  [arXiv:hep-th/0305159] \semi
  M.~R.~Douglas, I.~R.~Klebanov, D.~Kutasov, J.~Maldacena, E.~Martinec and N.~Seiberg,
  ``A new hat for the c = 1 matrix model,''
  arXiv:hep-th/0307195 \semi
  T.~Takayanagi and N.~Toumbas,
  ``A matrix model dual of type 0B string theory in two dimensions,''
  JHEP {\bf 0307}, 064 (2003)
  [arXiv:hep-th/0307083].
}

\lref\McGreevyEP{
  J.~McGreevy, J.~Teschner and H.~Verlinde,
  ``Classical and quantum D-branes in 2D string theory,''
  JHEP {\bf 0401}, 039 (2004)
  [arXiv:hep-th/0305194].
}

\lref\DiVecchiaNE{
  P.~Di Vecchia and A.~Liccardo,
  ``Gauge theories from D branes,''
  arXiv:hep-th/0307104.
}

\lref\klebstras{
  I.~R.~Klebanov and M.~J.~Strassler,
  ``Supergravity and a confining gauge theory: Duality cascades and
  chiSB-resolution of naked singularities,''
  JHEP {\bf 0008}, 052 (2000)
  [arXiv:hep-th/0007191].
}

\lref\polstras{
  J.~Polchinski and M.~J.~Strassler,
  ``The string dual of a confining four-dimensional gauge theory,''
  arXiv:hep-th/0003136.
}

\lref\maldanunez{
  J.~M.~Maldacena and C.~Nunez,
  ``Towards the large N limit of pure N = 1 super Yang Mills,''
  Phys.\ Rev.\ Lett.\  {\bf 86}, 588 (2001)
  [arXiv:hep-th/0008001].
}

\lref\maldalunin{
  O.~Lunin and J.~Maldacena,
  ``Deforming field theories with U(1) x U(1) global symmetry and their gravity
  duals,''
  arXiv:hep-th/0502086.
}

\lref\HananyIE{
  A.~Hanany and E.~Witten,
  ``Type IIB superstrings, BPS monopoles, and three-dimensional gauge dynamics,''
  Nucl.\ Phys.\ B {\bf 492}, 152 (1997)
  [arXiv:hep-th/9611230].
}

\lref\PolchinskiMT{
  J.~Polchinski,
  ``Dirichlet-Branes and Ramond-Ramond Charges,''
  Phys.\ Rev.\ Lett.\  {\bf 75}, 4724 (1995)
  [arXiv:hep-th/9510017].
}

\lref\MaldacenaRE{
  J.~M.~Maldacena,
  ``The large N limit of superconformal field theories and supergravity,''
  Adv.\ Theor.\ Math.\ Phys.\  {\bf 2}, 231 (1998)
  [Int.\ J.\ Theor.\ Phys.\  {\bf 38}, 1113 (1999)]
  [arXiv:hep-th/9711200].
}

\lref\EguchiTC{
T.~Eguchi and Y.~Sugawara,
``Modular invariance in superstring on Calabi-Yau n-fold with A-D-E
singularity,''
Nucl.\ Phys.\ B {\bf 577}, 3 (2000)
[arXiv:hep-th/0002100].
}

\lref\MurthyES{
S.~Murthy,
``Notes on non-critical superstrings in various dimensions,''
JHEP {\bf 0311}, 056 (2003)
[arXiv:hep-th/0305197].
}

\lref\KlebanovYA{
I.~R.~Klebanov and J.~M.~Maldacena,
``Superconformal gauge theories and non-critical superstrings,''
arXiv:hep-th/0409133.
}

\lref\KS{S. Kuperstein and J. Sonnenschein,
  ``Non-critical supergravity (d > 1) and holography,''
  JHEP {\bf 0407}, 049 (2004) [arXiv:hep-th/0403254].}

\lref\KazamaQP{
  Y.~Kazama and H.~Suzuki,
  ``New N=2 Superconformal Field Theories And Superstring Compactification,''
  Nucl.\ Phys.\ B {\bf 321}, 232 (1989).
}

\lref\ZamolodchikovAH{
A.~B.~Zamolodchikov and A.~B.~Zamolodchikov,
``Liouville field theory on a pseudosphere,''
arXiv:hep-th/0101152.
}

\lref\ElitzurPQ{
S.~Elitzur, A.~Giveon, D.~Kutasov, E.~Rabinovici and G.~Sarkissian,
``D-branes in the background of NS fivebranes,''
JHEP {\bf 0008}, 046 (2000)
[arXiv:hep-th/0005052].
}

\lref\BoucherBH{
W.~Boucher, D.~Friedan and A.~Kent,
``Determinant Formulae And Unitarity For The N=2 Superconformal Algebras In
Two-Dimensions Or Exact Results On String Compactification,''
Phys.\ Lett.\ B {\bf 172}, 316 (1986).
}

\lref\NamHU{
S.~k.~Nam,
``Superconformal And Super Kac-Moody Invariant Quantum Field Theories In
Two-Dimensions,''
Phys.\ Lett.\ B {\bf 187}, 340 (1987).
}

\lref\KiritsisRV{
E.~Kiritsis,
``Character Formulae And The Structure Of The Representations Of The N=1, N=2
Superconformal Algebras,''
Int.\ J.\ Mod.\ Phys.\ A {\bf 3}, 1871 (1988).
}

\lref\AharonyXN{
O.~Aharony, A.~Giveon and D.~Kutasov,
``LSZ in LST,''
Nucl.\ Phys.\ B {\bf 691}, 3 (2004)
[arXiv:hep-th/0404016].
}

\lref\Giveon{
  A.~Giveon, A.~Konechny, A.~Pakman and A.~Sever,
  ``Type 0 strings in a 2-d black hole,''
  JHEP {\bf 0310}, 025 (2003)
  [arXiv:hep-th/0309056].
}

\lref\WittenYR{
E.~Witten,
``On string theory and black holes,''
Phys.\ Rev.\ D {\bf 44}, 314 (1991).
}

\lref\DijkgraafBA{
R.~Dijkgraaf, H.~Verlinde and E.~Verlinde,
``String propagation in a black hole geometry,''
Nucl.\ Phys.\ B {\bf 371}, 269 (1992).
}

\lref\HoriAX{
K.~Hori and A.~Kapustin,
``Duality of the fermionic 2d black hole and N = 2 Liouville theory as  mirror
symmetry,''
JHEP {\bf 0108}, 045 (2001)
[arXiv:hep-th/0104202].
}

\lref\EguchiIK{
T.~Eguchi and Y.~Sugawara,
``Modular bootstrap for boundary N = 2 Liouville theory,''
JHEP {\bf 0401}, 025 (2004)
[arXiv:hep-th/0311141].
}

\lref\OoguriCK{
  H.~Ooguri, Y.~Oz and Z.~Yin,
  ``D-branes on Calabi-Yau spaces and their mirrors,''
  Nucl.\ Phys.\ B {\bf 477}, 407 (1996)
  [arXiv:hep-th/9606112].
}

\lref\KutasovPV{
D.~Kutasov,
``Some properties of (non)critical strings,''
arXiv:hep-th/9110041.
}

\lref\KutasovUA{
D.~Kutasov and N.~Seiberg,
``Noncritical Superstrings,''
Phys.\ Lett.\ B {\bf 251}, 67 (1990).
}

\lref\EguchiYI{
T.~Eguchi and Y.~Sugawara,
``SL(2,R)/U(1) supercoset and elliptic genera of non-compact Calabi-Yau
manifolds,''
JHEP {\bf 0405}, 014 (2004)
[arXiv:hep-th/0403193].
}

\lref\FotopoulosUT{
A.~Fotopoulos, V.~Niarchos and N.~Prezas,
``D-branes and extended characters in SL(2,R)/U(1),''
Nucl.\ Phys.\ B {\bf 710}, 309 (2005)
[arXiv:hep-th/0406017].
}

\lref\DiVecchiaRH{
P.~Di Vecchia and A.~Liccardo,
``D branes in string theory. I,''
NATO Adv.\ Study Inst.\ Ser.\ C.\ Math.\ Phys.\ Sci.\  {\bf 556}, 1 (2000)
[arXiv:hep-th/9912161].
}

\lref\DiVecchiaFX{
P.~Di Vecchia and A.~Liccardo,
``D-branes in string theory. II,''
arXiv:hep-th/9912275.
}

\lref\GaberdielJR{
M.~R.~Gaberdiel,
``Lectures on non-BPS Dirichlet branes,''
Class.\ Quant.\ Grav.\  {\bf 17}, 3483 (2000)
[arXiv:hep-th/0005029].
}

\lref\HananyTB{
  A.~Hanany and A.~Zaffaroni,
  ``On the realization of chiral four-dimensional gauge theories using
  branes,''
  JHEP {\bf 9805}, 001 (1998)
  [arXiv:hep-th/9801134].
}

\lref\GiveonPX{
A.~Giveon and D.~Kutasov,
``Little string theory in a double scaling limit,''
JHEP {\bf 9910}, 034 (1999)
[arXiv:hep-th/9909110].
}

\lref\FZZ{
V.~A.~Fateev, A.~B.~Zamolodchikov and Al.~B.~Zamolodchikov, unpublished.
}

\lref\AharonyUB{
O.~Aharony, M.~Berkooz, D.~Kutasov and N.~Seiberg,
``Linear dilatons, NS5-branes and holography,''
JHEP {\bf 9810}, 004 (1998)
[arXiv:hep-th/9808149].
}

\lref\GiveonZM{
A.~Giveon, D.~Kutasov and O.~Pelc,
``Holography for non-critical superstrings,''
JHEP {\bf 9910}, 035 (1999)
[arXiv:hep-th/9907178].
}

\lref\HikidaXU{
  Y.~Hikida and Y.~Sugawara,
  ``Superstring vacua of 4-dimensional pp-waves with enhanced  supersymmetry,''
  JHEP {\bf 0210}, 067 (2002)
  [arXiv:hep-th/0207124].
}

\lref\RecknagelSB{
  A.~Recknagel and V.~Schomerus,
  ``D-branes in Gepner models,''
  Nucl.\ Phys.\ B {\bf 531}, 185 (1998)
  [arXiv:hep-th/9712186].
}

\lref\GutperleHB{
  M.~Gutperle and Y.~Satoh,
  ``D-branes in Gepner models and supersymmetry,''
  Nucl.\ Phys.\ B {\bf 543}, 73 (1999)
  [arXiv:hep-th/9808080].
}

\lref\WittenSC{
  E.~Witten,
  ``Solutions of four-dimensional field theories via M-theory,''
  Nucl.\ Phys.\ B {\bf 500}, 3 (1997)
  [arXiv:hep-th/9703166].
}

\lref\ElitzurPQ{
S.~Elitzur, A.~Giveon, D.~Kutasov, E.~Rabinovici and G.~Sarkissian,
``D-branes in the background of NS fivebranes,''
JHEP {\bf 0008}, 046 (2000)
[arXiv:hep-th/0005052].
}

\lref\FateevIK{
V.~Fateev, A.~B.~Zamolodchikov and A.~B.~Zamolodchikov,
``Boundary Liouville field theory. I: Boundary state and boundary  two-point
function,''
arXiv:hep-th/0001012.
}

\lref\TeschnerMD{
J.~Teschner,
``Remarks on Liouville theory with boundary,''
arXiv:hep-th/0009138.
}

\lref\ZamolodchikovAH{
A.~B.~Zamolodchikov and A.~B.~Zamolodchikov,
``Liouville field theory on a pseudosphere,''
arXiv:hep-th/0101152.
}

\lref\RibaultSS{
S.~Ribault and V.~Schomerus,
``Branes in the 2-D black hole,''
JHEP {\bf 0402}, 019 (2004)
[arXiv:hep-th/0310024].
}

\lref\EguchiIK{
T.~Eguchi and Y.~Sugawara,
``Modular bootstrap for boundary N = 2 Liouville theory,''
JHEP {\bf 0401}, 025 (2004)
[arXiv:hep-th/0311141].
}

\lref\AhnTT{
C.~Ahn, M.~Stanishkov and M.~Yamamoto,
``One-point functions of N = 2 super-Liouville theory with boundary,''
Nucl.\ Phys.\ B {\bf 683}, 177 (2004)
[arXiv:hep-th/0311169].
}

\lref\IsraelJT{
D.~Israel, A.~Pakman and J.~Troost,
``D-branes in N = 2 Liouville theory and its mirror,''
arXiv:hep-th/0405259.
}

\lref\AhnQB{
C.~Ahn, M.~Stanishkov and M.~Yamamoto,
``ZZ-branes of N = 2 super-Liouville theory,''
JHEP {\bf 0407}, 057 (2004)
[arXiv:hep-th/0405274].
}

\lref\HosomichiPH{
K.~Hosomichi,
``N = 2 Liouville theory with boundary,''
arXiv:hep-th/0408172.
}

\lref\IsraelFN{
D.~Israel, A.~Pakman and J.~Troost,
``D-branes in little string theory,''
arXiv:hep-th/0502073.
}

\lref\BilalUH{
A.~Bilal and J.~L.~Gervais,
``New Critical Dimensions For String Theories,''
Nucl.\ Phys.\ B {\bf 284}, 397 (1987).
}

\lref\BilalIA{
A.~Bilal and J.~L.~Gervais,
``Modular Invariance For Closed Strings At The New Critical Dimensions,''
Phys.\ Lett.\ B {\bf 187}, 39 (1987).
}

\lref\WittenSC{
E.~Witten,
``Solutions of four-dimensional field theories via M-theory,''
Nucl.\ Phys.\ B {\bf 500}, 3 (1997)
[arXiv:hep-th/9703166].
}

\lref\BarsSR{
I.~Bars and K.~Sfetsos,
``Conformally exact metric and dilaton in string theory on curved
space-time,''
Phys.\ Rev.\ D {\bf 46}, 4510 (1992)
[arXiv:hep-th/9206006].
}

\lref\TseytlinMY{
A.~A.~Tseytlin,
``Conformal sigma models corresponding to gauged Wess-Zumino-Witten
theories,''
Nucl.\ Phys.\ B {\bf 411}, 509 (1994)
[arXiv:hep-th/9302083].
}

\lref\KutasovFG{
  D.~Kutasov, K.~Okuyama, J.~w.~Park, N.~Seiberg and D.~Shih,
  JHEP {\bf 0408}, 026 (2004)
  [arXiv:hep-th/0406030].
}

\lref\BertoliniGG{
  M.~Bertolini, P.~Di Vecchia, G.~Ferretti and R.~Marotta,
  Nucl.\ Phys.\ B {\bf 630}, 222 (2002)
  [arXiv:hep-th/0112187].
}

\lref\bianchi{
M.~Bianchi, G.~Pradisi and A.~Sagnotti, Nucl.\ Phys.\ B {\bf B376}, (1992)  365}

\lref\calnap{
 C.~G.~.~Callan, C.~Lovelace, C.~R.~Nappi and S.~A.~Yost,
 ``Adding Holes And Crosscaps To The Superstring,''
 Nucl.\ Phys.\ B {\bf 293}, 83 (1987).
}

\lref\FNP{A.~Fotopoulos, V.~Niarchos and N.~Prezas,
 ``D-branes and SQCD in Non-Critical Superstring Theory,''
 [arXiv:hep-th/0504010].}

\lref\diVec{
P.~Di Vecchia, M.~Frau, I.~Pesando, S.~Sciuto, A.~Lerda and R.~Russo,
``Classical p-branes from boundary state,''
Nucl.\ Phys.\ B {\bf 507}, 259 (1997)
[arXiv:hep-th/9707068].
}

\lref\eguchi{
T.~Eguchi and Y.~Sugawara,
``Modular bootstrap for boundary N = 2 Liouville theory,''
JHEP {\bf 0401}, 025 (2004)
[arXiv:hep-th/0311141].
}

\lref\gaberdiel{
M.~R.~Gaberdiel,
``Lectures on non-BPS Dirichlet branes,''
Class.\ Quant.\ Grav.\  {\bf 17}, 3483 (2000)
[arXiv:hep-th/0005029].
}

\lref\janbranes{
D.~Israel, A.~Pakman and J.~Troost,
``D-branes in N = 2 Liouville theory and its mirror,''
arXiv:hep-th/0405259.
}

\lref\kutsei{
D.~Kutasov and N.~Seiberg,
``Noncritical Superstrings,''
Phys.\ Lett.\ B {\bf 251} (1990) 67.
}

\lref\li{
 M.~Li,
 ``Boundary States of D-Branes and Dy-Strings,''
 Nucl.\ Phys.\ B {\bf 460}, 351 (1996)
 [arXiv:hep-th/9510161].
}

\lref\sameer{
S.~Murthy,
``Notes on non-critical superstrings in various dimensions,''
JHEP {\bf 0311}, 056 (2003)
[arXiv:hep-th/0305197].
}

\lref\ouyang{
 I.~R.~Klebanov, P.~Ouyang and E.~Witten,
 ``A gravity dual of the chiral anomaly,''
Phys.\ Rev.\ D {\bf 65}, 105007 (2002) [arXiv:hep-th/0202056].
}

\lref\polchinski{J.~Polchinski, ``String theory, Vol. I, II'', Cambridge University Press (1998).}

\lref\ribschom{
  S.~Ribault and V.~Schomerus,
  ``Branes in the 2-D black hole,''
  JHEP {\bf 0402}, 019 (2004)
  [arXiv:hep-th/0310024].
}

\lref\yost{
 S.~A.~Yost,
 ``Bosonized Superstring Boundary States And Partition Functions,''
 Nucl.\ Phys.\ B {\bf 321}, 629 (1989).
}

\lref\farkas{H.~M.~Farkas and I.~Kra, ``Theta constants, Riemann surfaces and the modular group'', Graduate studies in mathematics, Vol.37. Amer.\ Math.\ Soc.\ }

\lref\tatatheta{D.~Mumford, ``Tata Lectures on Theta''}

\lref\mizo{S.~Mizoguchi, ``Modular invariant critical superstrings on four-dimensional Minkowski space $\times$ two-dimensional black hole'', JHEP {\bf 0004} 14 (2000) [arXiv:hep-th/0003053].}

\lref\bilal{
A.~Bilal and J.~L.~Gervais, ``New critical dimensions for string theories'', Nucl.\ Phys.\ B {\bf 284}, 397 (1987). }

\lref\absteg{
M.~Abramowitz and I.~Stegun, ``Handbook of Mathematical Functions''.
}

\lref\gepner{
 D.~Gepner, ``Lectures On N=2 String Theory,''
PUPT-1121 {\it Lectures at Spring School on Superstrings, Trieste, Italy, Apr 3-14, 1989} }


\Title{\vbox{\baselineskip12pt\hbox{hep-th/0504079}
}}%
{\vbox{\centerline{D-branes in non-critical superstrings} 
\vskip8pt  
\centerline{and minimal super Yang-Mills in various dimensions.}
}}

{\vskip -20pt\baselineskip 14pt   
\centerline{Sujay K. Ashok$^{a,1}$  
\footnote{}{$^1$\tt ashok@physics.rutgers.edu},  
Sameer Murthy$^{b,2}$ \footnote{}{$^2$\tt smurthy@ictp.trieste.it}  
and  
Jan Troost$^{c,3}$ \footnote{}{$^3$\tt troost@lpt.ens.fr} }  
  
\bigskip  

\centerline{\sl $^a$ Tata Institute for Fundamental Research,}
\centerline{\sl  Homi Bhabha Road, Mumbai 400 005 India. }
\centerline{\sl and }
\centerline{\sl Department of Physics and Astronomy} 
\centerline{\sl Rutgers University, Piscataway, NJ 08855-0849, USA.}  
\smallskip  
\centerline{\sl $^b$Abdus Salam International Center for Theoretical Physics}  
\centerline{\sl Strada Costiera 11, Trieste, 34014, Italy.}  
\smallskip  
\centerline{\sl $^c$Laboratoire de Physique Th\'eorique}  
\centerline{\sl Ecole Normale Sup\'erieure${*}$ \footnote{}{$*$Unit{\'e} mixte  du
CNRS et de l'Ecole Normale Sup{\'e}rieure, UMR 8549. Preprint: LPTENS-05/15.}}
\centerline{\sl 24, Rue Lhomond, Paris 75005,  France.}
\bigskip\bigskip

\centerline{\bf Abstract}
We construct and analyze D-branes in superstring
theories in even dimensions less than ten. The backgrounds under study are supersymmetric 
$R^{d-1,1} \times SL(2,R)_k / U(1)$ 
where the level of the supercoset is  tuned such as to provide bona fide string theory backgrounds. 
We provide exact boundary states for D-branes that are localized at the tip of the cigar $SL(2,R)/U(1)$ supercoset  conformal field theory. We analyze the spectra of open 
strings on these D-branes and show explicitly that they are consistent with 
supersymmetry in $d=2,4$ and $6$. The low energy theory on the world-volume of the D-brane in each case is pure Yang-Mills theory with minimal supersymmetry. 
In the case with four macroscopic flat directions $d=4$, we realize an $\CN=1$
super Yang-Mills theory, and we interpret the backreaction for the dilaton as the running of the gauge
coupling, and study the relation between R-symmetry breaking in the gauge theory 
and the backreaction on the RR axion. 
\noindent   
}

\Date{July 2005}

\newsec{Introduction}
\subsec{Non-critical Superstrings and Holography}
It has been proven useful to study the physics of gauge theories using the
geometrical pictures and intuition provided by brane set-ups in string theory
(see e.g. \HananyIE).
One spectacular outcome of the study of D-branes and their associated
geometry \PolchinskiMT\ has been the impressive list of concrete examples of
holography \MaldacenaRE,
in which gravitational theories are dual to theories without a massless
spin two particle. 

In this paper, we concentrate on backgrounds of string theory with 
$d$ flat directions, supplemented with a cigar superconformal field theory
$R^{1,d-1}$ times $SL(2,R)/U(1)$ \refs{\kutsei, \EguchiTC , \mizo  ,\sameer }. 
The background can arise from taking
a double scaling limit of string theory near a singularity in a Calabi-Yau
manifold, or in the presence of NS5-branes \GiveonZM. It can also be thought of 
 as providing a $d$-dimensional
string theory background per se. 
By dialing the level $k$ the background becomes
critical. For even dimension $d$, the background comes equipped with
an $N=2$ superconformal worldsheet supersymmetry that can be used to
GSO project and that provides us with a target-space supersymmetric superstring theory.

These backgrounds and their D-branes are appropriate examples to further
study the interplay between holography and D-branes. Indeed, these 
solutions are of linear dilaton type and can be argued to interpolate
between two-dimensional string theories and their ten-dimensional cousins.
The dilaton gradient provided by the cigar conformal field theory 
takes values intermediate between the two-dimensional (strong) gradient
and the ten-dimensional (zero) gradient. This mechanism for achieving
criticality allows us to interpolate in the dimension of space-time. Since
two-dimensional (or generally low-dimensional) examples of holography seem to
be under more control than their ten-dimensional counterparts (see e.g. \mcgreevy\  and follow-ups)
it may be worthwhile to lay out the playground in  between. In the process,
 we should learn more about linear dilaton holography \AharonyUB .

\subsec{Gauge Theory Physics}
Constructing string duals to $\CN=1$ SYM
 theories has proven to be difficult. Previous approaches  \refs{\klebstras,
 \maldanunez} start from bulk theories with a larger number of
supersymmetries,
which are then broken through various mechanisms. 
An unwanted feature in these constructions is the existence of (extra) matter fields
(e.g. massive scalars and/or fermions) in the theory. When one goes to the deep 
infrared (where the extra  matter fields are absent), the supergravity 
backgrounds typically are not under control, either because of 
strong curvature \klebstras\ or strong coupling \maldanunez .

Non-critical superstrings seem to be free of most of these problems. The
target-space has reduced space-time supersymmetry, and the branes living in
them are likewise less supersymmetric. 
Thus, we need only carefully construct the bulk, and then the
 corresponding branes to study gauge theories with less 
supersymmetry. In this sense, the occurrence of gauge theories with less
supersymmetry is natural in the context of 
lower-dimensional superstring backgrounds. 
One must keep in mind however that the curvatures in these backgrounds 
are of string scale; gravity is not a priori a good approximation, and 
one necessarily has to work with the full sigma model which (after
backreaction) involves the 
difficult problem of dealing with background fluxes. 

In this paper, we present a boundary state description
of the branes in these backgrounds. Though it {\it is} a closed string
description in principle, the boundary states are 
in practice more useful to describe open string physics by
channel duality, and less so to compute the exact 
string background;  we only compute the linear
backreaction to the closed string background. 
Nevertheless, we consider this an important first step. 
This approach to $\CN=1$ gauge theories through lower-dimensional superstrings
may provide us with a new window to gauge theory physics.

\subsec{ Summary of Results}
In this paper, we present the exact conformal field theory description of
branes in lower-dimensional superstring theories, and compute the spectrum and
low energy theory on these branes for $d=2,4,6$. In the $d=4$ case, we present 
evidence that our closed string background is dual 
to a non-gravitational theory which in the IR flows arbitrarily close to $4d$
$\CN=1$ SYM, and in the UV is completed to a theory
which is asymptotically free.
In this sense, it is similar to the holographic descriptions in 
\polstras\ and \maldanunez \foot{Indeed the profile
of the closed string fields closely resembles that 
of the solutions of \maldanunez . }.

Furthermore, it is possible to understand instantons and the anomalous
 breaking  of the chiral $U(1)_R$ symmetry to $Z_{2N}$ along the lines 
 of \ouyang\ by studying the large
 distance 
  behavior of the Ramond-Ramond fields: the background value of the RR axion
 potential spontaneously breaks a $U(1)$ isometry of the solution.

\subsec{Organization}
In section 2 we briefly review the bulk physics of noncritical superstring theories. The boundary states that describe the D-branes that we concentrate on are presented in section
3. We analyze the spectrum encoded in the one-loop partition 
function in some detail and argue for the low-energy effective action
for these branes in section 4. We follow up by
laying bare the physics encoded in the one-point function of the boundary
states, in the case of $N=1$ SYM in four dimensions. 
In the conclusions we summarize our results and indicate possible
further developments. Various technicalities, and a generic proof of
the vanishing of open string partition functions in Gepner-like
(compact or non-compact) models are presented in the appendices.

\subsec{Note added in publication}
Very recently the interesting paper \FNP\ appeared with some overlap with our paper.
In particular, we note the overlap in the construction\foot{Following prior work in \eguchi.}  
of the open string spectrum for $d=4$. 
However, it is mostly usefully complementary in both subject matter and techniques.
In \FNP\ one finds an explicit analysis  of the relation to brane set-ups and
an analysis of 
flavor physics in this context. In our work, we focus on pure Yang-Mills,
providing 
many details of the  open string theory. We moreover compute properties of
the dual closed string background
using the boundary states.

\newsec{Superstrings in dimensions less than ten}
\seclab\bulk
In this section, we briefly review salient features of the 
closed string background in which we will embed D-branes 
in section 3. 
The closed string background we shall study is the type IIB $d$ dimensional
superstring \kutsei\ which 
consists of $d$-dimensional 
flat space tensored with a non-trivially curved space:
 $\IR^{d-1,1} \times SL(2,R)/U(1)$. The factor
$SL(2)/U(1)$ is a Kazama-Suzuki 
supersymmetric coset conformal field theory \KazamaQP , at (supersymmetric) level $k$. One can write an effective target space action for this coset as:
\eqn\cigarbackgnd{
ds^{2} = k\left(d\rho^{2}+ \tanh^{2}{\rho} d \theta^{2}\right) \, , \quad e^{\Phi} = {e^{\Phi_{0}} \over \cosh{\rho}}
}
This SCFT is known to have a mirror description as the 
$\CN=2$ Liouville theory. The level $k$ of the coset is 
tuned to make the total central charge $c=15$. The matter
worldsheet theory is tensored with the 
standard $\CN=2$ superconformal  ghosts of central charge $c=-15$,
such that the total
worldsheet central charge 
 vanishes. From the formula for the flat space and coset central charge,
we derive a relation between the dimension $d$ and the level $k$:
\eqn\cencharge{
c  = 15 = {3d\over 2} + 3 + {6\over k}.
}
For future reference we note that we have the following correspondences:
$$\eqalign{
\hbox{For}\quad d=6 &\quad k=2; \qquad \hbox{For}\quad d=4 \quad k=1; \cr
\hbox{For}\quad d=2 &\quad k={2\over3};\qquad 
\hbox{For}\quad d=0 \quad k={1\over2}.
}$$
For $d=8$ we obtain the familiar superstring in ten dimensional flat space, while for
$d=0$ we obtain a critical two-dimensional black hole background. The supercoset
 theory asymptotes to a $\CN=2$ linear dilaton with slope $Q=\sqrt{2 \over
 k}$. The short calculation above illustrates how the level of the coset (i.e. dilaton
gradient) allows us to interpolate in the dimension of space-time.

We will now discuss briefly the two-dimensional conformal field theories on the worldsheet.
In the $d+2$ dimensional theory, the free scalar fields $X^{\mu}$ parameterize the flat
space directions. Far from the tip of the cigar, the cigar can be approximated
by a
cylinder with a dilaton varying linearly along its length. The cylinder
directions will be labeled by the fields ($\rho,\theta)$.\foot{The capital 
letters $I,J=0,1..d-1,\rho,\theta$ label all the $d+2$ dimensions
in the theory while the Greek indices $\mu,\nu=0,1..d-1$ run over the flat
space directions only. }
For each worldsheet boson there is a corresponding worldsheet fermion.
In the flat directions, we have worldsheet fermions $\psi^{\mu}$,
while the fermions in the cigar directions are
named $\psi^{a}_{cig}$.

The superconformal field theory on $\IR^{d-1,1}$ and the cigar CFT are essentially decoupled: the
\CN=2 worldsheet currents are the 
sums of the respective currents of the two theories, and states are built in
the product state space of the two conformal field theories. The closed string vertex operators are the
operators on the cigar
 $\Phi^{j}_{m,\mb}$ multiplied with the vertex operators on $\IR^{d-1,1}$
\eqn\clstringops{
\ket{\CV(k)} = \ket{V_{X,\psi,gh}(k)} \otimes\, \Phi^{j}_{m,\mb}\ket{0}_{Cig} \,.
}
Here, $(j,m,\mb)$ label the primaries of the $SL(2)/U(1)$ supercoset. We
review the construction of these states in more detail in appendix B.
Here, we note that the quantum number $j$ governs the radial behavior of the
wavefunctions, asymptotically $\phi \sim e^{2 j \rho}$; the quantum numbers
$m,\mb$ are related to the momentum and winding around the cigar.  The bulk
superstring theories have an interesting physical spectrum that depends
strongly on the dimension -- this was analyzed in  \refs{\EguchiTC,  \mizo,
\sameer} to which we refer for details. Below we explicitly write down the
vertex operators
 necessary for the analysis of one point functions we carry out in section 5.

\subsec{Closed String Vertex Operators for $d=4$}

\item{\it a)} {\it The Graviton}

\medskip

In the NSNS sector, we shall first consider states in the $(-1, -1)$ picture with $m=\mb=0$. The worldsheet states which give us the second rank tensor in spacetime are:
\eqn\bulkNSNS{
\ket{\CV^{j \, IJ}(k)}= \psi^{I}_{-\half}\, \t \psi^{J}_{-\half}   \left[  \ket{0,k}_{X,\psi,NS}    \otimes\,    \Phi^{j}_{00} \ket{0}_{Cig,NS}  \otimes\, \ket{0}_{gh}  \right].
}
Physical states obey the condition $L_{0} - 1= \half + \half + \half k^{\mu} k_{\mu} - {j(j+1)\over k} -1=0$. We will be 
interested in the graviton modes which propagate in the radial direction of
the cigar. These have $k_{\mu} k^{\mu} = 0$ and are in the continuous
representation on the cigar $j=-\half+iP$. The on-shell condition becomes 
$P^{2} = - 1/4$.

\medskip

\item{\it b)} {\it The Tachyon}

\medskip

As mentioned in section 2, the non-trivial part of the closed string
background can be thought of as a supercoset or equivalently as an $\CN=2$
Liouville theory with winding condensate. It is immediate from the second
description (but it is also seen easily from the first) that there is a stable
scalar field (called the tachyon) in the spectrum with asymptotic winding
number one.
 This mode of the tachyon has the form in the $(-1,-1)$ picture
\eqn\windone{
\ket{\CV^{j}(k)}= \ket{0,k}_{X,\psi,NS}    \otimes\,    \Phi^{j}_{\half \half} \ket{0}_{Cig,NS}  \otimes\, \ket{0}_{gh} .
}
The mass shell condition for this mode is $L_{0} - 1 = \half+ \half k^{\mu} k_{\mu} + {1 \over 4} - {j(j+1)} -1 = 0 $. For $k_{\mu} k^{\mu}=0$, and $j= -\half + iP$, the on-shell state becomes  $P^{2}=0$. 

\medskip

\item{\it c)} {\it The Ramond-Ramond Axion}

\medskip

We are interested in the zero mode of the axion field and so we restrict to operators with $m=\mb=0$.  As we will see below, the calculation of the one point function on the disk forces the Ramond sector vertex operator to be in the $(-{3\over 2},-\half)$ picture.
The full vertex operator is obtained by tensoring with the ghost contributions and four dimensional spin fields. For propagating states with $m=\mb=0$ and $k^{\mu}k_{\mu}=0$, the on-shell condition is $L_0 -1 =  \left[{3\over 8} + 2\cdot {1\over 8}\right] + \left[P^2 + {1\over 4}+ {1\over 8}\right] - 1 = P^{2}=0$. 

The symmetry of the backreaction problem 
tells us that the components of the one form field strength along the flat
directions vanish: $\p_{\mu} \chi =0$. The remaining modes $\p_{\theta \pm
\rho} \chi$ are constructed \sameer\ from the spin fields such that there is
zero spin in the flat four directions and  the $U(1)$ $R$ charge of the
$\CN=2$ algebra on the cigar is {\bf $Q= \half$}. The same result holds for
the right moving sector. The vertex
 operator in the $(-\half, -\half)$ picture may now be written as
\eqn\fullRRop{
\ket{\CV_{-\half -\half}(k)}=S^{\a}\,\t S_{\a}\left[\ket{0,k}_{X,\psi} \otimes\Phi^{j}_{\half,\half} \ket{0}_{Cig}\otimes \ket{0}_{gh}\right]
}
where the $4$-dimensional spinorial index  $\a$ is contracted to get a scalar.

We note that the behavior of the graviton which had an effective mass in the
six dimensions is different from that of the tachyon and axion which 
are effectively massless. 
We shall see later that this difference manifests
 itself in the difference in the falloff rates in 
the weak coupling region of the backreaction onto these fields.

\subsec{General remarks about non-critical strings}

The string coupling at the tip of the cigar $g_{s}^{tip} = e^{\Phi_{0}}$ is a
 modulus of the theory and  is related to the parameter multiplying the
 Sine-Liouville interaction in the mirror description. Starting from the
 linear dilaton theory, one can obtain one or the other description by turning
 on the operator $\t \mu \Phi^{-1}_{00}$ or $\mu \Phi^{k \over 2}_{\half
 \half}$. The parameters are related as $(g_{s}^{tip})^{-2} = \left( {\mu
 \over k }\right)^{2 \over k} = \t \mu { \Gamma(1/k) \over \Gamma(1-1/k)}
 \equiv \nu^{-1}$.  All the bulk amplitudes of the theory depend on this
 parameter. This will also be true of the localized
 branes that we discuss in this paper. 

The various theories have a bosonic Poincare symmetry generated by the momenta and Lorentz rotations in flat space The theory with $d$ flat directions has $2^{d\over 2}$ left moving conserved supercharges. For $d\over 2$ even, there are two sets of conjugate spinors $\CS_{\a}, \t\CS_{\dot \a}$ and for $d \over 2$ odd, there are two sets of the same spinor  $\CS_{\a}, \t\CS_{\a}$. 
There are an equal number of  right moving supercharges, which for the type IIB theory obey exactly the same condition. There is another conserved charge, the $U(1)$ momentum around the cigar, which acts as an $R$-symmetry in the $d$-dimensional superalgebra. 
We have (in the case of even ${d\over 2}$, for example): 
\eqn\supalg{\eqalign{
\{ \CS_{a}, \bar{\CS}_{\dot \b} \} & = 2 \gamma^{\mu}_{\a \dot \b} P_{\mu} , \quad  \{ \T{\CS}_{a}, \T{\bar{\CS}}_{\dot \b} \}  = 2 \gamma^{\mu}_{\a \dot \b} P_{\mu} , \quad 
\{ \CS, \T \CS\}  = 0. \cr
[P^{\theta}, \CS_{\a}] & = \half \CS_{\a} , \quad [P^{\theta}, \bar{\CS}_{\dot \a}]  = - \half \bar{\CS}_{\dot \a}. \cr
[P^{\theta}, \T{\CS}_{\a}] & = \half \T \CS_{\a} , \quad [P^{\theta}, \T{\bar{\CS}}_{\dot \a}]  = - \half \T{\bar{\CS}}_{\dot \a}. \cr
}}

We have set up the discussion of lower-dimensional string theories without
referring explicitly to their ten-dimensional origins since we believe they
deserve study in their own right, and exhibit physics that are very
particular to their precise form (a simple example being the dimension
of space-time). However, it is often helpful to realize their roots in ten
dimensions. They arise from NS5-branes or singularities inside a
Calabi-Yau manifold, in a double scaling limit \GiveonZM\  \GiveonPX\ in which the string coupling is
taken to zero while keeping fixed the mass of the relevant non-abelian degrees
of freedom (leading to the Higgsed phase of little string theories). From the
perspective of the exact description of the near-horizon geometry in terms of
coset conformal field theories, one can lose dimensions of space-time by tuning the
value of the level of any number of $SU(2)/U(1)$ factors such that they become of central charge
zero. (This has its well-known analogue in the Landau-Ginzburg worldsheet description of
the strings near singularities in  Calabi-Yau manifolds embedded in weighted
projective spaces.) The analysis of the exact description of branes in NS5-brane backgrounds  \ElitzurPQ \eguchi \IsraelFN\ is thus technically close to the analysis that follows in section 3.

\newsec{The boundary state}
\seclab\boundarystates

In this section we review the ingredients that are necessary to construct
boundary states in the full lower-dimensional string theory that are
consistent with the bulk spectrum and the GSO projection \eguchi . To that end, 
we need aspects of boundary states in flat space, as well as 
boundary states in the supersymmetric cigar conformal 
field theory \refs{\ribschom ,\janbranes }.
We add to this a careful analysis of the Ramond ground state to complete the
construction of the full boundary state. We will then use the boundary states assembled 
in this section to analyze the physics of the spectrum as well as of
the one-point function in sections 4, 5 and 6.
We assemble some of the details of the set-up in appendix A.

\subsec{The terms and factors in the boundary state}
We first recall the different ingredients in the boundary state.
\ndt The branes we focus on are of the form:
\eqn\formofbranes{
\ket{\bB} = {T \over 2} \sum_{\alpha} \ket{B_{X,\psi}}_{\alpha} \otimes \ket{B}_{Cig\,, \alpha} \otimes \ket{B_{gh}}_{\alpha}\, ,
}
where the $\ket{B_{X,\psi}}$ refers to that part of the boundary state coming
from the 
flat space $\IR^{1,d-1}$ directions and $\ket{B}_{Cig}$ refers to a boundary
state 
in the cigar conformal field theory. The part of the boundary state $\ket{B_{gh}}$ in the
ghost sector is 
identical to the one constructed for $Dp$-branes in  ten dimensional
superstring theory \refs{\calnap , \yost }.
The sum denoted by $\alpha$ above will run over the periodicity (namely the NS-NS and R-R
sectors) 
and spin structures (which encode how the left and right fermions are glued
together for a 
given periodicity). 

The building blocks that constitute the boundary states are the Ishibashi
states which satisfy  
the gluing conditions for a fixed label $\alpha$. They  can be solved for separately in the
NS-NS and R-R 
sectors of the theory. 
In appendix A, we give a detailed construction of the D-branes that are
extended in all of the flat spacetime directions, and that are point-like in the
cigar directions.

We briefly recall the solution to the bosonic part of the conditions on the boundary
state:
\eqn\bosonsoln{
\ket{Bp_{X}} = \exp{\left[ -\sum_{n=1}^{\infty}{1\over n} \alpha_{-n}^{\mu} \eta_{\mu \nu}\t\alpha_{-n}^{\nu}\right]}\ket{0,k^{\mu}=0}.
}
The ket $\ket{0,k^{\mu}=0}$ denotes the vacuum of the worldsheet bosons. In the
fermionic sector, 
we need to account for the different periodicities and spin structures. 
The solution to the non-zero modes of the fermionic equations is
\eqn\fermionsoln{
\ket{Bp_{\psi}, \eta}_{NS/R} = \exp{\left[ i \eta \sum_{r >0}^{\infty} \psi_{-r}^{\mu} \eta_{\mu \nu}\t\psi_{-r}^{\nu}\right]}\ket{0, \eta}_{NS/R}.
}
Here $\ket{0, \eta}_{NS/R}$ is the fermionic vacuum. There is a unique NS
sector vacuum. However in the R sector, we also need to solve the zero mode fermionic
constraints in order to specify the vacuum. Since the total dimension of
space-time differs  from
ten, leading to a different Clifford algebra satisfied by the fermion
zeromodes, this calculation differs slightly from the usual one. We give the relevant
technical details in the appendix A.

\subsec{Assembling the full boundary state}
In order to describe the full boundary state, we must be more specific about
the sum over periodicities and spin structures in the boundary
 state \formofbranes . To construct a GSO invariant boundary state, we need to
sum over NS and R sectors, and then insert a projection operator
$(1+(-)^{F})(1+(-)^{\t F})$ in the type IIB superstring theory. 
The sum over the label $\alpha$ is 
then a sum over four terms, either NS or R and either with or without
the  insertion of the operator $(-)^{F}$.\foot{We note here that the raising operators in the
Ishibashi states do not change the relative $(-)^{F}$ between
 the left and right movers. In the NS sector, the vacuum and hence all the
boundary states that we have been considering have $(-)^{F}=(-)^{\t F}$. In
the R sector, the vacuum has to be chosen with a certain
 value of $(-)^{F+\T F}$, and all the states then retain that choice.  In
either case, the sum over
 eight terms thus reduces to four terms.}  This sum is equivalent to a sum
over NS and R-sector
and the two values of the spin structure $\eta$.\foot{In both the NS and R sector, the boundary state satisfies
$(-)^{F}\ket{\eta} = \pm \ket{-\eta}$, and this facilitates the solution of
the GSO projected state as $(\ket{\eta=+} \pm \ket{\eta=-})$. 
In the boundary state $\eta$ actually labels whether $(-)^{F}$ is present or not in the sum.} 
We index this sum by the label $\alpha=(NS, \T{NS}, R, {\T R})$. 

The boundary states of flat space to  be tensored with the cigar part are:
\eqn\flatblocks{
\ket{Bp_{X,\psi}}_{\a} = \ket{Bp_{X}} \otimes \ket{Bp_{\psi}}_{\a}
}
where the right hand side of the equation is given by equations
\bosonsoln\ and \fermionsoln. Finally, in the cigar sector, we read off from
\refs{\EguchiIK ,\janbranes} (see also \refs{\AhnQB , \HosomichiPH})
the expression for
the boundary state corresponding to a point-like brane on the cigar: 
\eqn\cigarbranes{
\ket{D0}_{Cig\, \a} = \sum_{j,m,\mb}  \Psi^{j, \, \alpha}_{m,\mb} \; \Phi^{j}_{m,\mb} \ket{0}_{\a}.
 }
 As explained in Appendix B, one can separate the supercoset into a bosonic part and free fermions. 
The wavefunctions in the different sectors are\foot{We choose $u=1$ in the
notation of that paper, i.e. the D0-brane on the cigar with only the extended trivial
representation  in the open string channel.} determined by the purely bosonic part of the CFT:
\eqn\wavefn{\eqalign{
 \Psi^{j,NS/R}_{m_{bos} \mb_{bos}} & =  k^{-\half} (-1)^{m_{bos}+\mb_{bos}\over k}\,\delta_{m_{bos},\mb_{bos}}\, \nu^{1/2+j} {\Gamma(-j+m_{bos}) \Gamma(-j-m_{bos}) \over \Gamma(-2j-1) \Gamma(1-{1+2j \over k})} 
 \cr \Psi^{j,\, \T NS/ \T R}_{m_{bos}\mb_{bos}} & =  i^{m_{bos} + \mb_{bos} \over k} \Psi^{j,\, NS}_{m_{bos}\mb_{bos}} \cr 
}}
The full GSO projected boundary state can now be written down using the
factors
\flatblocks\ and \cigarbranes:
\eqn\finbdryst{
\ket{{\bf Bp}} =\sum_{\alpha} \ket{Bp_{X, \psi}}_{\alpha} \otimes \ket{D0}_{Cig,\alpha}\otimes \ket{B_{gh}}; \qquad \alpha=NS, \T{NS}, R, {\T R}.
} 
The explicit form of the ghost part of the boundary state will not play a role
in the computations that
 follow and so have not been written out.
We have used the notation for the tensor product which
is  correct only 
in the NS sector. The GSO projection ties together the R sector vacua of the
factors, and the equation should be read as representing
 the tensor product of
the raising operators 
acting on the total vacuum.

A Cardy type check can be performed on these D-branes. 
It consists of a combination of Cardy checks which
have already been performed on the individual factors that comprise the
full boundary state. The calculation is therefore a combination of the
usual Cardy check performed on D-branes in flat space, and the D-branes
of the cigar conformal field theory. Note that for the branes that are localized on the
cigar,
the spectrum in the open string channel is discrete, indeed allowing for a
standard Cardy check in terms of the demand that open string
degeneracies
are positive integers -- this is not always the case in non-rational
conformal field theories where open string partition functions can depend
on continuous quantum numbers, and where volume divergences can spoil this
approach tailored on rational conformal field theories. However, for the localized
branes on which we concentrated in this paper, no such complication arises.
The Cardy check is thus straightforward.

\newsec{The open string theory on the branes}
In this section we discuss the physics associated to the D-branes we
constructed above. In particular, we study in some detail the low-energy
spectrum
for the open strings living on the D-branes, and the low-energy effective
action that describes their dynamics.
The D-branes presented in the previous section break half of the bulk space-time supersymmetry, which
will be indicated by the presences of massless fermions (goldstinos). The
 other half of the bulk supersymmetry is linearly realized on the brane -- in
 appendix C, we exhibit the explicit
 form of the supercharges preserved by the D-brane in the four-dimensional case $d=4$, using a worldsheet analysis.

\ndt In the following we present for each D-brane, as a function of the
dimension $d$: 
\item{1.} The spectrum of excitations on the D-brane and information on the
low energy limit of the worldvolume theory. In every case, the theory is a pure
gauge theory with minimal supersymmetry, and the spectrum consists of gauge
bosons and gauginos which are the realization of the goldstinos.
\item{2.} The exact form of the full partition function and a proof that it
vanishes for $d=4,6$, consistent  with supersymmetry. The case $d=2$ is a little subtle, because of the potential existence of unpaired fermion zero modes, on which we shall comment briefly. 
\item{3.} A considerably more general proof of supersymmetry following arguments
used for supersymmetric bulk partition functions \gepner\ is presented in appendix D. 

\vskip 0.2cm

We first summarize the parts of the analysis that are common to all space-time dimensions.
In all dimensions $d$, the partition function for the branes filling the flat space $\IR^{d-1,1}$ is
given by the following sum over sectors labeled by $\alpha$:
\eqn\pbrnpfn{
Z_{Dp}(t) = \half \left(Z_{Dp}^{NS}(t) - Z_{Dp}^{\T{NS}}(t) - Z_{Dp}^{R}(t) - Z_{Dp}^{\T R}(t) \right) , 
}
where the individual terms are given by the expressions \refs{\GaberdielJR , \eguchi}:
\eqn\pbrnpieces{\eqalign{
Z_{Dp}^{NS}(t)  & = V_{d}  \int {d^d k\over (2\pi)^d} e^{-2\pi t k^2}
\left({\Theta_{00} (it)
 \over \eta^{3}(it)} \right)^{d-2 \over 2} \times {\Theta_{00}(it) \over \eta^{3}(it)} \sum_{s \in \IZ + \half}  {1 \over 1+q^{s}} \left(q^{s^{2}-s \over k} - q^{s^{2}+s \over k} \right). \cr
Z^{\T{NS}}_{Dp}(t) & = V_{d} \int {d^d k\over (2\pi)^d} e^{-2\pi t k^2}
\left({\Theta_{01} (it) 
\over \eta^{3}(it)}\right)^{d-2 \over 2} \times {\Theta_{01}(it) \over
\eta^{3}(it)} 
\sum_{s \in \IZ + \half}  {(-1)^{s-\half} \over 1-q^{s}} \left(q^{s^{2}-s \over k} + q^{s^{2}+s \over k}\right). \cr
Z_{Dp}^{R}(t) & = V_{d} \int {d^d k\over (2\pi)^d} e^{-2\pi t k^2}
\left({\Theta_{10} (it) \over \eta^{3}(it)}\right)^{d-2 \over 2} \times
{\Theta_{10}(it) \over \eta^{3}(it)} \sum_{s \in \IZ}  {1 \over 1+q^{s}}
 \left(q^{s^{2}-s \over k} - q^{s^{2}+s \over k}\right).\cr
Z_{Dp}^{\T R}(t) & = V_{d} \int {d^d k\over (2\pi)^d} e^{-2\pi t k^2}
\left({\Theta_{11} (it) \over \eta^{3}(it)}\right)^{d-2 \over 2} \times
{\Theta_{11}(it) \over \eta^{3}(it)} \sum_{s \in \IZ}  {(-1)^{s} \over 1-
q^{s}} 
\left(q^{s^{2}-s \over k} + q^{s^{2}+s \over k}\right).\cr
}}

The worldvolume theory on the branes has a bosonic Poincare symmetry in the $d$ flat directions. 
We show in appendix C that exactly half of the supercharges \supalg\ are preserved by the brane. These are of the form $\CS_{\a}+ \T{\CS}_{\a} \equiv \CS_{\a}^{bdry}$, $\bar{\CS}_{\dot \a}+ \T{\bar{\CS}}_{\dot \a} \equiv \bar{\CS}_{\dot \a}^{bdry}$. In the free theory, the $U(1)_{R}$ symmetry is preserved. The superalgebra is of the form:
\eqn\supalgbndry{\eqalign{
\{ \CS^{bdry}_{\a}, \bar{\CS}^{bdry}_{\dot \b} \}  &= 2 \gamma^{\mu}_{\a \dot \b} P_{\mu}  \cr
[P^{\theta}, \CS^{bdry}_{\a}]  = \half \CS^{bdry}_{\a} ,& \quad [P^{\theta}, \bar{\CS}^{bdry}_{\dot \a}]  = - \half \bar{\CS}^{bdry}_{\dot \a} \cr
}}

\ndt In all the open string theories, we can write down the following massless states:
\eqn\lightmodesgen{
\epsilon_{\nu} \psi^{\nu}_{-\half} \ket{k_{\mu},NS}, \qquad u^{\alpha}\ket{k_{\mu},\Sigma_{\alpha}, R}.
} 
Here, $\Sigma_{\alpha}$ is the  spin field on the worldsheet which 
transforms under $Spin(d)$. These states are  BRST invariant on the worldsheet
when $k^{\mu}k_{\mu}=0$, and when the polarizations $\epsilon^{\mu}$ and
$u^{\alpha}$ obey the conditions 
$k^{\mu} \epsilon_{\mu}=0, \; \epsilon^{\mu} \equiv \epsilon^{\mu} + k^{\mu}$ and $k^{\mu} \gamma_{\mu} u = 0$. 
The physical interpretation of these modes are clear as a gauge boson and a gaugino in each of the cases.
Since the other modes on the brane have masses of order $\alpha'^{-\half}$,
 there is a sensible low energy limit
 in which one can write down a low energy action for these massless modes. 
In the following, we will turn to the individual cases $d=4,6,2$ and discuss
some features of the theories which are typical to the bulk theory in which
they
are embedded. We start out with the four-dimensional theory.

\subsec{$d=4$ and $\CN=1$ SYM}

\ndt {\it The low energy theory}

Writing the partition function \pbrnpieces\ in the $NS$ sector as
\eqn\partnfnNS{
Z^{NS}(q) = \int {d^4 k\over (2\pi)^4} q^{k^2}\, A^{NS}(q)\,, 
}
the masses of the excitations in sector $\alpha$ are obtained by expanding
$A^{\a}$ in powers of $q=e^{-2\pi t}$. The coefficients in the expansion give 
the degeneracy of states with a given mass. We find:
\eqn\NSfourdexp{
A^{NS}(q) = q^{-{1\over 2}} + 2  + 4 q^{1\over 2} + 12 q + \ldots 
}
The first state with negative conformal dimension is the $NS$ sector vacuum
and will be projected out by the GSO projection. 
The lowest lying physical states in the $NS$ sector are therefore two massless gauge bosons from the first excited level. In the $R$-sector, one finds the expansion
\eqn\Rfourdexp{
\half A^{R}(q) = \left( \sum_{s\in \IZ_+} q^{s^2-s}\, (1-q^s)^2\right) \cdot 2 \prod_m {(1+q^m)^4\over (1-q^m)^4} = 2 + 12q+ 52 q^2 + \dots  
}
The partition function in the twisted $R$ sector vanishes, and thus, the (GSO projected)
Ramond sector gives rise to two physical massless fermionic states in spacetime. 
 
To summarize, we  see that the modes on the brane with $k_{\mu}^{2}=0$ are two physical states each from the NS / R sector.  Analyzing \lightmodesgen\ for this case, we find that the spectrum consists of a massless gauge boson  $A_{\mu}$ and a massless gaugino $\lambda_{\a}$ transforming in the ${\bf 2}$ of $Spin(4)$, which are the physical degrees of freedom corresponding to the $\CN=1$ SYM multiplet in four dimensions. 

In appendix C, we have shown that the D-brane boundary state \finbdryst\ preserves $\CN=1$ supersymmetry in $d=4$.  We deduce that the low energy ($\alpha' \to 0$) effective action is indeed that of pure super Yang-Mills theory 
\eqn\neqoneym{
S_{YM}= {1 \over g^{2}_{YM}} \int d^{4}x \; \Tr \left({1 \over 4} F^{2} + \bar{\lambda} \p \lambda \right). 
}

\ndt {\it Full partition function}

\ndt Using the identities in Appendix E, we can rewrite  \pbrnpfn\ as:
\eqn\fullpart{\eqalign{
Z_{D3}(\tau) =   \int {d^4 k\over (2\pi)^4} e^{2\pi i \tau k^2} {1 \over \eta^6(\tau)}\,
&\left[ \Theta^{2}_{00}(\tau) \left( - \Theta_{10}(2\tau) + e^{-{i \pi \tau \over 2}}  \Theta_{00}(2\tau) \right) \right. \cr
&- \left. \Theta^{2}_{01}(\tau) \left(  \Theta_{10}(2\tau) + e^{-{i \pi \tau \over 2}}  \Theta_{00}(2\tau) \right)
\right. \cr 
&-\left.  \Theta^{2}_{10}(\tau) \left( - \Theta_{00}(2\tau) + e^{-{i \pi \tau \over 2}}  \Theta_{10}(2\tau) \right) \right] 
}}
Now we use the following identities of theta functions \refs{\farkas, \mizo, \bilal},  
\eqn\thetaid{\eqalign{
\Theta^{2}_{00}(\tau) & = \Theta^{2}_{00}(2 \tau) + \Theta^{2}_{10}(2 \tau), \cr 
\Theta^{2}_{01}(\tau) & = \Theta^{2}_{00}(2 \tau) - \Theta^{2}_{10}(2 \tau), \cr 
\Theta^{2}_{10}(\tau) & = 2 \Theta_{00}(2 \tau) \Theta_{10}(2 \tau), \cr 
}}
and plugging into \fullpart , we find that 
\eqn\partvan{
Z_{D3}(\tau) = 0,
}
consistent with supersymmetry.

\subsec{$d=6$, $\CN=(0,1)$ SYM}

\ndt {\it Low energy theory}

\ndt From \pbrnpieces, using manipulations similar to the $d=4$ case, we find 
\eqn\expsixd{\eqalign{
A^{NS}(q) &= q^{-\half} + 4 + 13\,q^{\half}+40\,q+106\,q^{3\over 2}+256\,q^2+\ldots \cr
A^{{\T NS}}(q) &= q^{-\half} - 4 + 13\,q^{\half}-40\,q+106\,q^{3\over 2}-256\,q^2+\ldots \cr
\half A^{R}(q) &= 4 + 40\,q+256\,q^2+\ldots\,  \cr 
}}

The $D5$-branes preserve eight supercharges, and the low energy theory is then determined to be the  $(0,1)$ supersymmetric Yang-Mills theory in six dimensions. 

\vskip 0.2 cm

\ndt {\it Full partition function}

\ndt As sketched in Appendix E, one can rewrite the partition function as:
\eqn\sixdpfnparts{\eqalign{
Z^{NS}& = {\Theta_{00}^{4}(\tau)\over \eta^9(\tau)}  \sum_{r=0}^{\infty} (-1)^{r}  \left(q^{-\half (r-\half)^{2}}-  q^{-\half (r+\half)^{2}}   \right) \cr
Z^{\T NS}  & = {\Theta_{01}^{4}(\tau)\over \eta^9(\tau)}  \sum_{r=0}^{\infty}
(-1)^{r}  \left(q^{-\half (r-\half)^{2}}-  q^{-\half (r+\half)^{2}}   \right) \cr
Z^{R} & = {\Theta_{10}^{4}(\tau)\over \eta^9(\tau)}  \sum_{r=0}^{\infty}
(-1)^{r}  \left(q^{-\half (r-\half)^{2}}-  q^{-\half (r+\half)^{2}}   \right) \cr
Z^{\T R} & = 0 .\cr
}}

Using the commonly encountered Jacobi identity, we again get the result that the
partition function vanishes. 
\eqn\fullpartsixd{
Z_{D5}(\tau)=0.
}

\subsec{$d=2$ and $\CN=(0,2)$ SYM}

\ndt {\it Low energy theory}

\ndt For $d=2$, we find the $q$-expansion of the partition functions to be
\eqn\exptwod{\eqalign{
A^{NS} (q) &= q^{-\half} + 0 + q^{\half}+2\,q+5\,q^{3\over 2}+6\,q^2+8\,q^{5\over 2}+14\,q^3 +\ldots \cr
A^{\T{NS}} (q) &= q^{-\half} + 0 + q^{\half}-2\,q+5\,q^{3\over 2}-6\,q^2+8\,q^{5\over 2}-14\,q^3 +\ldots \cr
A^{R} (q) &= 2(1 + 2\,q+6\,q^2+14\,q^3+\ldots\,.) \cr
A^{\T R}(q) & = -2.
}}

By the now familiar argument of low energy spectrum, and exact supersymmetry, we find that in this case, the low energy theory on the D-strings is $d=2$, $\CN=(0,2)$ SYM. As this case has new features, we present below the full partition function. 

\vskip0.2cm

\ndt {\it Full partition function}

\ndt We have:
\eqn\twodpart{\eqalign{
Z^{NS} & =  {\Theta_{00}(\tau)\over \eta^3(\tau)} \sum_{s \in \IZ + \half}  {1
\over 1+q^{s}} 
\left(q^{3(s^{2}-s) \over 2} - q^{3(s^{2}+s) \over 2} \right)   \cr
Z^{\T NS} & =  {\Theta_{01}(\tau)\over \eta^3(\tau)}  \sum_{s \in \IZ + \half}
{(-1)^{s-\half}
 \over 1-q^{s}} \left( q^{3(s^{2}-s) \over 2} + q^{3(s^{2}+s) \over 2} \right)  \cr
Z^{R} & =  {\Theta_{10}(\tau)\over \eta^3(\tau)}  \sum_{s \in \IZ}  {1 \over
1+q^{s}} 
\left(q^{3(s^{2}-s) \over 2} - q^{3(s^{2}+s) \over 2}  \right)  \cr
Z^{\T R} & = - 2 \cr
}}

At this point, we would like to make a couple of

\ndt {\it Comments on fermion zero modes and vanishing of partition function:}

\item{1.} Note here that in $d=2$, we find two fermionic 
modes with $L_{0}=0$, both having $(-1)^{F}=-1$.  $Tr_{R}(-1)^{F}$ does not vanish due to the presence
of poles for two terms that come with the same sign. 
The usual free fermion degeneracy in the R sector is lifted in the coset for very particular states. 
This is seen as the fact that an operator $G_0^{\pm}$ annihilates a particular
state, 
instead of giving rise to a degenerate state, or as the presence of a pole in the partition function. 
Whenever the coset is combined with SCFT's with fermion zero modes (like for
$d=4,6$), we will not notice this subtlety
(since the extra excitations re-introduce the degeneracy, and make for the
vanishing of the total twisted R-sector partition function). 
 
\item{2.} It is still true that the partition function vanishes and two parallel D-strings do not feel any force. However, the fermion zero modes which are projected out need to be understood better, and may lead to interesting physics.

\newsec{Backreaction on the NSNS closed string background}
\seclab\onepoint 

In this section, we concentrate on the case of $d=4$ and
 compute the backreaction on the cigar background by $N$
 $D3-$branes. To first order, this is given by
 the appropriately
 transformed one-point function of closed string  fields in the presence of
 the boundary state \finbdryst\ multiplied by a factor of $N$. In particular, we shall compute the shift in the background
 value of the fields in the gravity multiplet, and the tachyon multiplet. We assume that the dilaton (the trace of the second rank tensor) couples  to the kinetic term in the gauge theory and interpret the profile of the dilaton in the radial direction of the cigar as the running gauge theory coupling (See also e.g.  \refs{\WittenSC ,
 \DiVecchiaNE , \BertoliniGG ,\IsraelFN } for similar phenomena in related contexts).

Indeed, earlier work on holography in asymptotically linear dilaton backgrounds 
 leads us to expect that closed string dynamics
 in the background of $N$ $D3-$branes in the six dimensional theory is
 dual to an open string theory which flows at 
low energies ($\rho \to 0$) arbitrarily close to the flow of pure $\CN=1$ SYM.
 Our first interest therefore is the
 backreaction on the dilaton near the tip of the cigar 
where one might expect agreement with the familiar logarithmic running of the
 coupling in the four dimensional supersymmetric gauge theory.

In order to get a ``geometric interpretation'', we use the following strategy:
We use as an input the exact form of the 
 one point functions and reflection amplitudes. As a tool during the
computation,
we approximate the exact closed string vertex operators with
their
mini-superspace approximate wavefunction. 
Even though we work in a strongly curved string background, where the
curvature is of order the string length, we can have some faith in our
calculations. First of all, the mini-superspace approximation will not be as
bad as it looks, since it is known that for $\CN=2$ supersymmetric coset
models the background metric and dilaton do not receive further curvature
corrections \refs{\BarsSR , \TseytlinMY}. Moreover, at appropriate intuitively
understandable junctions in the calculation, we replace semi-classical
approximations by their
 known exact counterparts.

\subsec{Bulk graviton}

The notation is as discussed in Section $2$ and the Appendices: the primary
vertex operators in the cigar SCFT are denoted
$\Phi^{j,n,\nb}_{m_{bos}\mb_{bos}}$. Their overlap with the boundary state is
called $\Psi^{j,n,\nb}_{m_{bos}\mb_{bos}}$, as in \wavefn. The minisuperspace
field configuration for 
these operators are denoted $\phi^{j}_{m}$.

The asymptotic behavior of the closed string field perturbation \bulkNSNS\  in
the presence of the D-branes \finbdryst\ is given (in  momentum space) by the
amplitude with the insertion of a 
closed string propagator (we are omitting to explicitly show that $n=\nb=0$ to avoid cluttering the equations):
\eqn\onept{
{1\over N} \T h^{IJ}(k^{\mu},P) \equiv \bra{\CV^{j \, IJ}(k^{\mu})} D_{cl} \ket{{\bf B3}} =
 \eta^{IJ} {\delta^{4}(k^{\mu})
 \over \half k_{\mu}k^{\mu} - j(j+1) } \Psi^{j,NS}_{00}(P)\,,
}
where $\CV$ is the vertex operator for the graviton introduced in section 2. To get the profile in position space, we fold this with the solution of the Laplacian in the six-dimensional 
background, which factorizes as $V_{k^{\mu},P}(x^{\mu}, \rho) = e^{ik_{\mu}X^{\mu}}
 \phi_{0}(\rho,P)$. 
The delta functions in the flat directions reduce the expression to a one dimensional integral:
\eqn\oneptpos{
{1 \over N} h^{IJ}(x^{\mu},\rho) =  \eta^{IJ} \int_{0}^{\infty} dP  \left( {\phi_{0}^{NS}(\rho,P) \over P^2+a^2} \right) \Psi^{NS}(P).
}
with $a=1/2$.
\ndt The exact one point function of the operator $\Phi^{-\half+iP}_{00}$ is \wavefn\ denoted:
\eqn\oneptP{
\Psi^{-\half+iP, NS}_{00} \equiv \Psi^{grav}(P) = \nu^{iP} {\Gamma(\half-iP)^{2} \over (\Gamma(-2iP)\Gamma(1-2iP))} \,. 
}
\ndt The delta-function normalized minisuperspace field $\phi^{-\half-iP,
 NS}_{0}$ is\foot{Note that in \oneptpos , the field used to implement the "Fourier
 transform" to get the profile in position space is the complex conjugate of the field
 whose one point function we compute in \oneptP .}  \ribschom :
\eqn\minisupspfield{
\phi^{-\half-iP, NS}_{0}(\rho) \equiv  \phi_{0}^{grav}(\rho,P) 
= - {\Gamma(\half+iP)^{2} \over \Gamma(2iP)} F\left(\half-iP, \half+iP; 1; -\sinh^{2}{\rho}\right).
}
where $F(a,b;c;z)$ is the hypergeometric function $_2F_1$. The exact reflection amplitude 
\eqn\refl{
R^{grav}(P)  \equiv { \Gamma(2iP)  \Gamma(\half-iP)^{2} \Gamma(1+2iP) \nu^{iP}
 \over  
\Gamma(-2iP) \Gamma(\half+iP)^{2} \Gamma(1-2iP) \nu^{-iP} } 
}
is obtained by putting $m_{bos}=\mb_{bos}=0$ in the expression for $R(P)$ in Appendix B; it obeys $R^{grav}(P) R^{grav}(-P)  =1$. Using this, the closed string field configuration can be written as:
\eqn\asympfield{\eqalign{
\phi_0^{grav}(\rho,P) & =  \left[ (\sinh{\rho})^{2iP-1} F\left(\half-iP, \half-iP; 1-2iP; -{1\over \sinh^{2}{\rho}}\right) \right. \cr
& \quad \left. + R^{grav}(-P) (\sinh{\rho})^{-2iP-1} F\left(\half+iP, \half+iP; 1+2iP; -{1\over \sinh^{2}{\rho}}\right)  \right] \,.
}}
In \asympfield, we have written the minisuperspace solution in terms of the basis of incoming and outgoing modes propagating on the cigar, which is better suited to the asymptotic description. If we rewrite the solution \minisupspfield\ in the above asymptotic basis by using the connection formula (Eg:
 $(15.3.7)$ of \absteg ), and replace the semiclassical reflection amplitude by the quantum one, we recover  \asympfield .

The expression for the linear backreaction on the graviton field is now:
\eqn\intindetail{\eqalign{
{1\over N} h^{IJ}(x^{\mu},\rho)  & =  \eta^{IJ}  \int_{0}^{\infty} dP {1\over P^{2} +a^{2}} \times 
\left[ \nu^{iP} {\Gamma(\half-iP)^{2} \over (\Gamma(-2iP)\Gamma(1-2iP))} \right. \cr
& \qquad \quad \left. (\sinh{\rho})^{2iP-1} F\left(\half-iP, \half-iP; 1-2iP;
-{1\over \sinh^{2}{\rho}}\right) \right] 
 + [P \leftrightarrow -P]
\cr
& =  \eta^{IJ} \int_{-\infty}^{\infty} dP {1\over P^{2} +a^{2}} \nu^{iP}
 {\Gamma(\half-iP)^{2} \over (\Gamma(-2iP)\Gamma(1-2iP))}
\times \cr
& \qquad \quad (\sinh{\rho})^{2iP-1} F\left(\half- iP, \half - iP; 1- 2iP; -{1\over \sinh^{2}{\rho}}\right) \cr
& \equiv \eta^{IJ}   I_{grav} \cr
}}

In fact the above integral needs a more precise contour prescription for it to
be  well-defined\foot{We would like to thank
Justin David and Edi Gava for emphasizing this important point.}. 
First of all, we note that if we had taken only the
semi-classical one-point function and reflection amplitude, the above
integral (without the factor $\Gamma(1-2iP)$ in the denominator) would allow
for closing the contour in the $P$ upper half-plane, leading to the evaluation
of the integral as the residue of the pole on the positive imaginary
axis\foot{Alternatively, one can compute the integral making use of a Schwinger parametrization
of the integral, and a fundamental integral representation of the
hypergeometric function.}.
For the above approximation to the exact result, a contour prescription will
be more subtle, since the behaviour of the extra $\Gamma$ function factor
does not allow for the naive semi-classical contour. Although
this subtlety is important,
we believe that a good approximation to the exact result is given by
evaluating the above integral at the pole in the propagator on the positive
imaginary axis.
(One can dictate a corresponding contour prescription.)
We will indeed see in the following in various instances that this
prescription captures a lot of the expected physics. The caveat that
we described will be present in further backreaction computations as well.

%

 The poles of the integrand are only at $P= \pm i a$, and we prescribed to
pick up the pole on the positive imaginary axis,
$P =  + ia$, $a>0$.
We have, with $a=\half$:
\eqn\intone{\eqalign{
I_{grav} & =  {2 \pi i \over 2 i a} \nu^{-a} (\sinh{\rho})^{-2a-1} F \left(\half + a,\half + a; 1+2a; - {1 \over \sinh^{2}\rho} \right) \cr
& = 2 \pi \nu^{-\half} (\sinh{\rho})^{-2} F\left(1,1;2; - {1 \over \sinh^{2}\rho} \right) \cr
& =   2 \pi \nu^{-\half} \log\left( 1+ {1 \over \sinh^{2}\rho} \right) \,.
}}
After multiplying by $N$, we get the expression for the graviton field:
\eqn\finalgrav{
h^{IJ}=  \eta^{IJ} 2 \pi N \nu^{-\half} \log\left( 1+ {1 \over \sinh^{2}\rho} \right) \,.
}
\ndt In the two limits of large and small radial distance, we find that
\eqn\gravitonasymp{\eqalign{
 h^{IJ}(\rho) & \longrightarrow \eta^{IJ} \  \left[  2 \pi N \nu^{-\half} e^{-2 \rho} \right] \qquad \hbox{as} \qquad \rho \to \infty \qquad \hbox{and}\cr
&  \longrightarrow \eta^{IJ}\  \left[ - 4 \pi N \nu^{-\half} \log{\rho}\right] \qquad\hbox{as} \qquad \rho \to 0 \,. 
}}

\ndt {\it Comments:}
\item{1.} We would like to point out here that the ``Fourier Transform'' we performed in order to convert the momentum space one-point function into the position space profile used only the {\it continuous series} $j=-\half + iP$. However, our result \intone\ is simply the profile of an on-shell mode in the {\it discrete series} with asymptotic behavior  $e^{-2\rho}$. The graviton mode with polarization in the cigar directions is precisely the interaction operator in the worldsheet theory  which is the first correction from the cylinder towards the cigar.\foot{The metric on the cigar \cigarbackgnd\ asymptotically looks like $ds^{2} = d\rho^{2} + d \theta^{2} + e^{-2 \rho} d\theta^{2}$.} 
\item{2.} This mode is normalizable at the weak coupling end. In this respect, our result is similar to the ones by \KutasovFG , in that the the localized branes sources the normalizable mode on the cigar.\foot{Our branes can be thought of as the analog of the ZZ branes of Liouville theory.}

\subsec{Relation to gauge theory}

The following is a short note on understanding the physics of the calculation above from the point of view of the holographic theory. The cigar background had a metric which was asymptotically flat and a dilaton which behaved as $\Phi(\rho)= - \log \cosh{\rho}$. The one-point function calculation tells us that  in the presence of the D-branes, the fields \bulkNSNS\ whose behaviour in the radial direction is $e^{\Phi}$ shift from the background value by  \intindetail , \intone . The trace of this second rank tensor gives us the change in the dilaton:
\eqn\dilpert{
\delta (e^{\Phi}) \sim  2 \pi N \nu^{-\half} \log\left( 1+ {1 \over \sinh^{2}\rho} \right).
}
To connect with the gauge theory on the D-branes, we use the expansion of this equation near the tip of the cigar where we get:
\eqn\dilpertagain{
\delta (e^{-\Phi})  \sim 4 \pi N \nu^{-\half} \log{\rho}. 
}
This can be understood semiclassically -- notice that the region near the tip cigar behaves like flat space\foot{Note that this is a space of  size string scale -- there is an overall factor of  $\alpha'$ which multiplies the metric.} with a constant dilaton.  In such a region, a pointlike source has a propagator which is logarithmic in two dimensions. If we use a Born-Infeld type action for the world-volume theory on the D-brane using the background closed string fields, 
\eqn\biact{
S_{D3} = \tau_{3} \int d^{4}x \, \left( e^{-\Phi} Tr F^{2}+ ...\right) 
}
and add this as a source to the closed string equations of motion, we recover \dilpert . Here $\tau_{3}$ is a dimensionful quantity entering the tension of the brane.

From the Born-Infeld action \biact , it is clear that $e^{-\Phi}$ acts as the coupling constant $g_{YM}^{-2}$ of the gauge theory. Also, in a putative holographic duality between $\CN=1$ SYM and the closed string theory in the background of these D-branes, it is reasonable to expect that the closed string field $e^{-\Phi}$ couples to the operator $Tr F^{2}$ in the action. 
Putting all of these facts together, we get:
\eqn\ymrun{
{1 \over g^{2}_{YM} } - {1 \over g^{2}_{YM,0}} \sim  N \log( \rho /\Lambda)
}

\ndt {\it Comments;}
\item{1.} We see here that the radial coordinate on the cigar $\rho$ plays the role of the scale in the gauge theory. The constant $\Lambda$, the strong coupling scale in the gauge theory is related to where the RG flow in the IR theory is matched to the full stringy UV complete theory. 
\item{2.} There is a factor of $\nu^{-\half}$ present in \finalgrav . We shall see in the next section that this should be interpreted as the renormalized string coupling at the tip $(g_{s,ren}^{tip})^{-1}$. This renormalization is a redefinition of the zero mode of the dilaton. In the gauge theory, such a redefinition would be a change in the strong coupling scale $\Lambda$.

\subsec{Bulk tachyon winding mode}

This state has $m=\mbar=\half$, and possesses the following reflection amplitude (with $k=1$):
\eqn\reftach{
R^{tach}(P) = \nu^{2iP} { \Gamma(2iP)   \Gamma(1+2iP)  \Gamma(1-iP) \Gamma(-iP ) 
 \over  
\Gamma(-2iP)  \Gamma(1-2iP)  \Gamma(1+iP) \Gamma(iP) }. 
}
The (generalized) wavefunction for this mode is 
\eqn\wavefntach{\eqalign{
\phi^{tach}_{P}(\rho) &=  \cosh \rho \; e^{\pm i \sqrt{1 \over 2}(\theta -\t \theta)} \; {\Gamma(1+iP) \Gamma(iP) \over \Gamma(1+2iP) \Gamma(2iP)}  F(1+iP,1-iP,1,-\sinh^{2} \rho ), \cr
& =  \cosh \rho  \; e^{\pm i \sqrt{1 \over 2}(\theta -\t \theta)} \; (\sinh{\rho})^{-2iP-2} F\left(1+iP, 1+iP; 1+2iP;-{1\over \sinh^{2}{\rho}}\right) +  \cr
& \qquad  \qquad R^{tach}(-P) \cosh \rho (\sinh{\rho})^{2iP-2} 
 F\left(1-iP, 1-iP; 1-2iP;-{1\over \sinh^{2}{\rho}}\right)   \,.
}}
In the second line, we rewrote the wavefunction in variables suited to the
asymptotic region consistent with the above reflection amplitude as in the
graviton case. We can now see easily that the asymptotic behavior is
$\Phi^{tach}(\rho) \equiv T(\rho) = \mu e^{\pm i {1 \over \sqrt{2}} (\theta
-\t \theta) - \rho} $. This behavior is on the edge of the Seiberg window of
operators non-normalizable at the weak coupling end. In this respect, the
$\CN=2$ Liouville theory with $k=1$ is similar to the $c=1$ bosonic Liouville
theory, and
 we can extend the understanding gained in that case \McGreevyEP\ to this one.

First, we note that objects in the theory are singular in the limit $k \to 1$,
 and we regularize as $k = 1 +  \epsilon$. In order to keep quantities like
 the two and three point functions in the bulk theory finite, we need to keep
 $\t \mu { \Gamma(1/k) \over \Gamma(1-1/k)} \equiv \nu^{-1}$ finite. Using the
 relation between the mirror parameters, this means that the bare $\CN=2$
 Liouville interaction  diverges.  To see what this implies, let us look at
 the full wavefunction of the tachyon winding mode including the
 reflected piece. The reflection amplitude for the mode 
in the action $j = {k \over 2}$ has the value $R=-1$. The asymptotic behavior is (keeping only the leading behavior): 
\eqn\tachaftref{\eqalign{
\Phi^{tach}(\rho) \equiv T^{phys}(\rho) & = \mu \; \lim_{\epsilon \to 0} e^{\pm i \sqrt{1+\epsilon \over 2}(\theta -\t \theta)}  \left( e^{-(1+\epsilon)\rho} + R(1-\epsilon)  e^{-(1-\epsilon)\rho} \right) \cr 
& = -(\mu \epsilon)\, \rho \, e^{\pm i {1 \over \sqrt{2}}(\theta -\t \theta)}  e^{-\rho} \equiv  \mu_{ren} \, \rho \, e^{\pm i {1 \over \sqrt{2}} (\theta -\t \theta) -\rho}  \cr
}}

{\it Comments:}
\item{1.} It is clear that we should keep the quantity $\mu_{ren} $ defined above finite and the relations between the various parameters $(g_{s,ren}^{tip})^{-2} = \mu_{ren}^{2} = \t \mu_{ren} \equiv \nu^{-1}$ where all the quantities are now finite and tunable.
\item{2.} Keeping track of all the terms in the above computation tells us that the full tachyon winding mode has also a normalizable piece which behaves as $\Phi \sim \mu_{ren} \log \mu_{ren} e^{\pm i {1 \over \sqrt{2}}(\theta -\t \theta)-\rho}$. We will see below that this is the mode that is sourced by the brane.

\subsec{Backreaction on the tachyon winding mode}

The one point function of this mode on our brane is given by:
\eqn\onpttach{
\Psi^{tach}(P) = - \nu^{iP} {\Gamma(1-iP) \Gamma(-iP) \over \Gamma(1-2iP) \Gamma(-2iP)}  
}
We can collect the above pieces as before to get the expression for the backreaction:
\eqn\backreactach{\eqalign{
\delta T (x^{\mu},\rho) & =   e^{\pm i \sqrt{1 \over 2}(\theta -\t \theta)} \; \int_{0}^{\infty} dP {1\over P^{2}} \times \left[ \nu^{iP} {\Gamma(1-iP) \Gamma(-iP) \over \Gamma(1-2iP) \Gamma(-2iP)} \right. \cr
& \qquad \quad \left. \cosh{\rho}  \; ( \sinh{\rho})^{2iP-2} F\left(1-iP, 1-iP; 1-2iP;
-{1\over \sinh^{2}{\rho}}\right) \right] 
 + [P \leftrightarrow -P]
\cr
& = e^{\pm i \sqrt{1 \over 2}(\theta -\t \theta)} \; \int_{-\infty}^{\infty} dP {1\over P^{2}} \nu^{iP}
{\Gamma(1-iP) \Gamma(-iP) \over \Gamma(1-2iP) \Gamma(-2iP)}
\times \cr
& \qquad  \quad \cosh{\rho} \; (\sinh{\rho})^{2iP-2} F\left(1-iP, 1-iP; 1-2iP;
-{1\over \sinh^{2}{\rho}}\right)  \cr
& \equiv e^{\pm i \sqrt{1 \over 2}(\theta -\t \theta)} \;  I_{tach} \cr
}}
We can compute the integral using the principal value
prescription\foot{For the subtleties involved in this procedure: see the
discussion following equation \intindetail\ .}:
\eqn\tachintone{\eqalign{
I_{tach} & = {1 \over \epsilon} \cosh{\rho} \; (\sinh{\rho})^{-2} F \left(1,1,1; - {1 \over \sinh^{2}\rho} \right) \cr
& = {1 \over \epsilon} (\cosh{\rho})^{-1} ; \cr
& \longrightarrow {1 \over \epsilon} e^{-\rho},\; \rho \to \infty; \cr
& \longrightarrow {1 \over \epsilon}  ,\; \rho \to 0;
}}
From the discussion of the bulk tachyon field above, it is clear now that the one-point function for the bare tachyon field\foot{Note that we have done the above backreaction calculation by using the expressions in the theory for general $k$ {\it without} renormalizing the parameters.} must diverge, and must be interpreted as $\delta T \sim (\delta \mu) e^{-\rho}$. The physical statement to be inferred from the above is:
\eqn\rentachback{
\delta T^{phys}(x^{\mu}, \rho) = \mu_{ren} e^{\pm i {1 \over \sqrt  2}(\theta -\t \theta)} \;  (\cosh{\rho})^{-1} \; .
}

%

\newsec{Backreaction on the RR fields}

We shall now repeat the analysis for the Ramond-Ramond fields which the $Dp$-brane source.  We shall focus on the case of the $D3$-brane which is charged under the dual of the axion field. 
The quantum numbers of the RR scalar has been discussed earlier in section 2. The reflection amplitude is  given by 
\eqn\RRreflection{
R^R(P) = \nu^{2iP} {\Gamma(2iP)\over \Gamma(-2iP)} {\Gamma(1-iP)\Gamma(-iP)\over \Gamma(1+iP)\Gamma(iP)} 
{\Gamma(1+2iP)\over \Gamma(1-2iP)} \,.
}
We now compute the profile of the RR field in spacetime following the procedure outlined in the NSNS case. We need to compute the overlap of the RR vertex operator with the boundary state in Appendix A in the $(-{3\over 2},-\half)$ picture  \refs{\bianchi ,\diVec}:
\eqn\RRoneptmom{
{1 \over N} \CA^{{\dot \a}\b}(k,P) = \bra{\CV^{{\dot \a}\b}_{-{3\over 2},-\half}(k)}D_{cl}\ket{{\bf B3}} = {\delta^4(k^{\mu})\over P^2}\,
\Psi^{j}_{\half,\half}\CN^{{\dot \alpha}\beta}\,.
}
where $j=-\half+iP$, and $\CN^{{\dot \a}\b}$ is the result of the zero mode overlap between the Ramond sector ground state of the boundary state and the spin fields. 
In this picture, we actually compute the background value of the gauge potential \diVec ,   by taking the trace of \RRoneptmom\ with the appropriate $\Gamma$-matrices:
$$
{1 \over N} A_{\mu_1\ldots\mu_n}(k) = \Tr(\CA(k)\,C\,\Gamma_{\mu_1}\ldots\Gamma_{\mu_n})\,.
$$
In position space, we thus get the profile of the gauge field to be
\eqn\RRoneptpos{
{1 \over N} A_{\mu_1\ldots\mu_n}(\rho) = \int_{0}^{\infty} {dP\over P^2}\, \phi_{P}^{R}(\rho)\,  \Psi_{P}^{R}\,\Tr(\CN\, C\, \Gamma_{\mu_1}\ldots\Gamma_{\mu_n})\,.
}
where $\Psi_P^R$ is given by \wavefn\  with $m_{bos}=\mb_{bos} = \half$
$$
\Psi_P^{R} \equiv \Psi^{j}_{\half,\half} = \nu^{iP}{\Gamma(1-iP)\Gamma(-iP)\over \Gamma(-2 iP)\Gamma(1-2iP)} \,.
$$ 
The Ramond sector of our boundary state \finbdryst\  tells us that the only non-zero gauge potential is $A_{0123}$. 

It now remains to obtain the solution to the Laplace equation $\phi_P^{R}(\rho)$ that implements the generalized Fourier transform. For modes with $m=\bar{m}$, this solution should have, in the semiclassical limit, the correct reflection amplitude obtained from the coset algebra. The required wavefunction in the minisuperspace approximation is given by
 \eqn\correctRRprofile{
\phi_{P}^{R}(\rho) \equiv \phi_P^{R,j=-\half-iP} = \cosh{\rho} \; {\Gamma(1+iP)\Gamma(iP)\over \Gamma(2iP)} 
F(1-iP,1+iP;1;-\sinh^2\rho) 
}
which has an asymptotic expansion:
\eqn\RRlaplacesoln{\eqalign{
\phi_P^{R}(\rho) &= \cosh{\rho} \; \left[(\sinh\rho)^{-2+2iP}\, F(1-iP,1-iP;1-2iP;-{1\over \sinh^2\rho}) \right. \cr
& \left. + R^R(-P)\,(\sinh\rho)^{-2-2iP}\, F(1+iP,1+iP;1+2iP,-{1\over \sinh^2\rho}) \right]\,,
}}
where the quantum reflection amplitude is given by \RRreflection. Indeed, as for the NS-NS case, we can use the connection formula for the hypergeometric function to check that it reproduces the classical reflection amplitude in the Ramond-Ramond sector \janbranes\ with
 $k\rightarrow \infty$. 

We can now repeat the analysis of the NS-NS sector. Substituting the above expressions in \RRoneptpos\ , we get 
\eqn\ramondint{\eqalign{
& {1 \over N} g_{s} A_{0123}(\rho) \cr
 &= \int_{-\infty}^{\infty} {dP\over P^2} 
 \nu^{iP}{ \cosh{\rho}\over (\sinh\rho)^{2-2iP}} {\Gamma(1-iP)\Gamma(-iP)\over\Gamma(-2iP)\Gamma(1-2iP)} F(1-iP,1-iP;1-2iP;-{1\over \sinh^2\rho}) \cr 
}}
This integral is divergent, with a behavior ${1 \over \epsilon} \cosh^{-1}\rho$ with a suitable regulator. To understand this, let us write the vertex operator in the $(-{3\over 2}, -\half)$ picture \diVec . The operator looks like 
\eqn\italRRvert{
W^{RR}(k) =  \CA_{\dot \a \b} (k) \CV^{\dot \a}_{-3/2}(k) \CV^{\b}_{-1/2}(k)+ i \CF_{\dot \a \dot \b} (k) \CV^{\dot \a}_{-3/2}(k) \CV^{\dot \b}_{-3/2}(k) \; \pb \t c \; \t \xi .
}
Let us write the vertex operators asymptotically with $\rho$ dependence $e^{(p- 1)\rho}$. 
BRST invariance of the vertex then implies $p\CF(p)=0$ and $p\CA(p) = \CF(p)$. In position space, these are $g_{s} d* g_{s}^{-1}  (g_{s}\CF) = 0$ and $g_{s} d g_{s}^{-1} \CA = g_{s} \CF$. 

This shows us the meaning of the divergence -- for a constant field strength, we have $\CA(p \to 0) = \CF(p \to 0)/p$, and in position space, this will translate to $\CA = \rho e^{-\rho}$. To keep all our calculations finite, it is also clear what to do: compute the field strength from the beginning:
\eqn\RRfstrint{\eqalign{
& {1 \over N} F_{0123\rho}(\rho) \equiv  I_R(\rho) \cr
&=  \p_{\rho} g_{s}^{-1} \int_{-\infty}^{\infty} {dP\over P^2} 
 \nu^{iP} { \cosh{\rho} \over (\sinh\rho)^{2-2iP}} {\Gamma(1-iP)\Gamma(-iP)\over\Gamma(-2iP)\Gamma(1-2iP)} F(1-iP,1-iP;1-2iP;-{1\over \sinh^2\rho}) \cr 
& = \int_{-\infty}^{\infty} {dP\over P^2} 
 \nu^{iP} \p_{\rho}\left[ {\cosh^{2}{\rho} \over (\sinh\rho)^{2-2iP}} {\Gamma(1-iP)\Gamma(-iP)\over\Gamma(-2iP)\Gamma(1-2iP)} F(1-iP,1-iP;1-2iP;-{1\over \sinh^2\rho}) \right] \cr 
 & = \int_{-\infty}^{\infty} {dP\over P^2} 
 \nu^{iP} (\sinh\rho)^{2iP} {\Gamma(1-iP)\Gamma(-iP)\over\Gamma(-2iP)\Gamma(1-2iP)}  \cr
& \qquad \qquad  \left[
\left( 2 \coth{\rho} + (-2 +iP) \coth^{3} \rho \right) F(1-iP,1-iP;1-2iP;-{1\over \sinh^2\rho}) \right. \cr
&\qquad \qquad + \left. 2 \coth^{3}\rho {1 \over \sinh^{2}\rho} {(1-iP)^{2} \over (1-2iP)}  F(2-iP,2-iP;2-2iP;-{1\over \sinh^2\rho}) \right]. \cr
}}
We use again the principal value of the integral and evaluating the pole as $P \to 0$, we get:
\eqn\rrans{\eqalign{
I_R & = {1 \over P} \left[ \left( 2 \coth{\rho} + (-2 +iP) \coth^{3} \rho \right) \tanh^{2} \rho + 2 \coth^{3}\rho {1 \over \sinh^{2}\rho} \tanh^{4} \rho \right] \cr
& = {1 \over P}  iP \coth^{3} \rho \tanh^{2} \rho \cr
& = \coth \rho. \cr
}}
Dualizing, we get $\chi \sim N \theta$. We shall fix the coefficient in the next section. For now, we 
note that we can integrate to get an expression for the (non-normalizable) vertex operator for the potential $g_{s} A_{0123}(\rho) = (\cosh \rho)^{-1} \log \sinh \rho $.

\subsec{A short note on instantons and the chiral $U(1)_{R}$ symmetry breaking.}

The $U(1)_{R}$ symmetry of the Super Yang-Mills theory is realized in the string theory dual as the conserved $U(1)$ momentum around the cigar. Adding the $D3$-branes at the tip sources a constant RR axion field strength, and the axion field hence depends linearly on the angular coordinate  with a coefficient proportional to $N$ which can be determined by channel duality of the annulus amplitude. We shall proceed to fix this coefficient using electric-magnetic duality of the action that describes the massless RR fields.

In the ten dimensional superstring, we know that the shift symmetry of the RR axion is non-perturbatively broken. This is also the case in the non-critical superstring theories, where the shift is realized as translation around the angular direction of the cigar $\theta$. To test this, as usual, we consider a D-instanton in the theory which is charged under this axion field. At all orders in perturbation theory, the zero mode of $\chi$ is a modulus, but this mode multiplies the action of a D-instanton, and for the string theory path integral to be well-defined even after summing over instanton configurations, we deduce that the zero mode is only defined upto periodic identifications. 

The type $IIB$ theory contains odd dimensional D-branes with a Chern-Simons coupling $\mu_{p} \int C_{p+1}$ for $p=-1,1,3,5$ where the RR potentials $C_{p}$ are canonically normalized. The action at tree level has a symmetry which exchanges electric and magnetic states under the various gauge potentials. The Dirac quantization condition then implies $\mu_{p}\mu_{2-p}=2\pi$. 

In the presence of $N$ $D3$-branes, we have $\chi = \mu_{3} N {\theta \over 2\pi}$. The Chern-Simons coupling of the D-instanton is then $\delta S_{D(-1)} =  \mu_{-1} \mu_{3} N  {\theta  \over 2\pi} = N \theta$; it follows that the geometric $U(1)$  isometry of the cigar is broken to $Z_{N}$.  Let us remind ourselves that the normalization of the $U(1)_{R}$ charge  \supalgbndry\ was defined such that the fermions in spacetime have a half-integer charge. This normalization is sensible from the geometric point of view as smooth boundary conditions at the tip of the cigar enforce antiperiodicity of the fermions. On the other hand, the normalization of the $U(1)_{R}$ current in the gauge theory is such that a rotation of $2 \pi $ gives the gluini a phase of unity. This makes it clear that $ \theta_{cig} = 2 \theta_{SYM} $. 

Putting the above facts together tells us that the modification of the closed string background shows a breaking of the  chiral $U(1)_{R}$ symmetry of the theory to ${\bf Z}_{2N}$ as expected. It is not clear from our construction how the chiral symmetry is broken further to $Z_{2}$. This is expected to involve the exact form of the axion field in the deep IR which is beyond the scope of this work.

\newsec{Conclusions}
We have taken an exact conformal field theory approach towards the
construction of interesting gauge theory physics in lower-dimensional superstring
theory backgrounds. The construction and explicit analysis of the full
open string spectrum was done using known boundary conformal
field theory results for the free scalar conformal field theory and 
the conformal field theory of the cigar $SL(2,R)/U(1)$. We concentrated
on the branes that are localized at the tip of the cigar, and analyzed
them in different target space dimensions. For $d=4$, we argued that the
low-energy spectrum and effective action are those of $N=1$ super Yang-Mills
theory.

Supersymmetry of the open string spectrum was shown directly in the
open string channel, in several complementary ways: technically and precisely, using 
non-trivial theta-function identities, and conceptually (and very generically), following the
techniques applied previously to supersymmetric bulk compactifications. The
conceptual proof applies to generic compact and non-compact Gepner models.

We analyzed some of the information encoded in the exact boundary states. 
We made a first analysis of the resulting backreaction on the
closed string background and the physics of the gauge theories that is
reflected
in it, e.g. the logarithmic running of the coupling
constant of $N=1$ super Yang-Mills theory, and the breaking of the $U(1)_R$
symmetry to ${\bf Z}_{2N}$.\foot{It would be desirable to investigate the further breaking to $Z_2$.}

Clearly, there is room for further analysis of the closed string backreaction.
In particular, one would like to go beyond the linear approximation 
for the pure $N=1$ super Yang-Mills theory. Since the 
compactification is at string scale, an exact conformal field theory approach
to this problem would be most convincing -- however, supersymmetry may
validate a low-energy approach. In particular, it would be desirable to have
an analytic solution
 to the six dimensional non-critical supergravity (along the lines of
\refs{\KS ,\KlebanovYA}) 
that corresponds to these D-branes. 

It is interesting to include flavors in the construction presented here 
\refs{\KS ,\KlebanovYA , \FNP}, to for instance 
compare the relative normalizations of the running
of the gauge coupling. Moreover,
it will be interesting to study chiral matter in the $\CN=1$ gauge
theory, following the techniques for obtaining chiral matter 
in brane set-ups (see e.g. \HananyTB\ and references therein).
 This should allow for a splitting of the 
multiplets analyzed in \refs{\ElitzurPQ , \FNP}.

A closer analysis of the gauge theory physics, for every individual even
dimension $d$ is equally desirable. We believe that the economical brane construction and tools provided in this paper
may serve a further analysis well.
Finally, we may hope that the exact conformal field theory treatment of this background
allows for a continuous interpolation between the more familiar 
gauge theory physics and the highly stringy physics at the typical length 
scale of the cigar (i.e. the string scale). We already saw an example of the 
convincing simplicity of this interpolation in the analysis of the linear
backreaction on the dilaton -- one may hope that this gives us a privileged
window on little string theory and holography at the string scale.

\newsec{Acknowledgments}
We would like to thank Bobby Acharya, Costas Bachas, Matteo Bertolini, Justin
David, Eleonora Dell'Aquila, Angelos Fotopoulos, Edi Gava, Dan Israel, Sunil
Mukhi, Vasilis Niarchos, Ari Pakman, Nikolaos Prezas and Ashoke Sen for
discussions on related topics and useful correspondence. 
S.A. would like to thank the kind hospitality
of the Harish-Chandra Research Institute, Allahabad, 
where part of this research was done. We would also like
 to thank the developers of Skype softphone that facilitated
 free inter-continental communication.

\appendix{A}{A few details of the construction of the full boundary states}

In the following, we will construct
D-branes that are 
extended in all of the flat spacetime directions, and that are point-like in the
cigar directions.\foot{We will discuss some of the physics related to
other choices of boundary states in due course.} 
The notion of being point-like is of course a semiclassical statement. The defining feature of the branes we shall study is that they are BPS (B branes), preserve the momentum around the cigar and localized near the tip. 

We will denote the boundary state by $\ket{Bp}$ with $p=d-1$. In the
$\IR^{1,d-1}$ directions, 
we have the worldsheet equations 
\eqn\neumann{\eqalign{
\p_{\tau} X^{\mu} (\s,0) \ket{Bp_{X,\psi}} & = 0, \cr
(\psi^{\mu} - i \eta {\t \psi}^{\mu}) \ket{Bp_{X,\psi}} & = 0, \quad \mu=0,.,p=d-1. 
}}
We will follow the conventions that the left-movers are holomorphic and the
right movers (indicated
by variables with tildes) are anti-holomorphic. In terms of the worldsheet modes, the equations become
\eqn\wsmodes{\eqalign{
(\alpha^{\mu}_{n}+ {\t \alpha}^{\mu}_{-n} )\ket{Bp_{X,\psi}}   & =0,  \cr
(\psi^{\mu}_{r} - i \eta {\t \psi}^{\mu}_{-r} )\ket{Bp_{X,\psi}}  & =0, \quad \mu=0,.,p\quad \forall\ n. 
}}
The equations
 \neumann\ and \wsmodes\ are written in worldsheet coordinates suited to the
closed string channel. 
Later on we rewrite these conditions in terms of the open string variables to derive
the supersymmetries
 that are left unbroken by the D-brane.
 
The zero mode of the bosonic oscillators is the momentum 
$k^{\mu}=\half(\alpha^{\mu}_{0}+\t \alpha^{\mu}_{0})$. In the fermionic
sector, 
$r \in \IZ + \half$ in the NS sector and $r \in \IZ$ in the R sector. 
The variable $\eta = \pm$ denotes the spin structure related to the $\IZ_{2}$ automorphism 
of the gauged $\CN=1$ algebra implemented by the map $G \to \eta G$. In the R sector, there are
fermion
 zero modes, and in that sector $\eta$ indicates the choice of the eigenvalue
of the operator $(-)^{F}$ acting on the ground state. 

In the cigar part, the conditions on the boundary state can be written down
in terms 
of the current algebra \OoguriCK \janbranes . The fermionic part involves the conditions
like those in equation \wsmodes\ for the fermions $\psi_{cig}^{\pm}$. The same spin structure $\eta$ as the flat
space part
 must be imposed on the fermionic modes of the cigar.\foot{This means that the fermions on the cigar also obey Neumann boundary conditions as shown below in the zero mode equation for the fermions. 
 We thank Angelos Fotopoulos, Vasilis Niarchos and Nikolaos Prezas for
correcting us on this point.}

There is another $\IZ_{2}$ ambiguity in the  choice relating the left
 and right moving
 $\CN=2$ currents which tells us whether we have $A$ or $B$ branes.  We shall be
 interested in $B$-type 
boundary conditions $J^{R}= -\bar{J}^{R}, \, G^{\pm}=i \eta \t G^{\pm}$ which
 leads to branes 
of the type \neumann\  extended along the flat directions and localized on the
 cigar. This leads
 also to  Neumann boundary conditions on the angular direction of the cigar
 \janbranes\ . These imply
 that the one point functions have a delta function in the momentum $n$ around the
 cigar.\foot{We note that the conventions of \EguchiIK \ are opposite to ours (the BPS
 localized D-branes in IIB 
theory are the A branes in \eguchi ) because they consider the theory on the
 T-dual circle with a
 condensate of momentum. Indeed, in the exact conformal field theory, the
 choice
of A- or B-type boundary condition is arbitrary, since they are related by an 
automorphism of the $\CN=2$ superconformal algebra. 
It is only after we specify a specific semi-classical picture,
 i.e. give a geometrical interpretation to the $\CN=2$ currents and the branes in the conformal field
 theory that the distinction between A-type and B-type branes becomes
 meaningful.} In the conventions 
introduced earlier, this means that $m=\mb$.

\subsec{Ramond ground state}
We need to solve for the zero mode factor of the boundary state
that solves the conditions in equation \wsmodes .
The R sector ground state of the theory also includes 
the solution of the zero modes equations of the two fermions of the coset.
The fact that the worldsheet fermions are free facilitates this analysis. 
We can treat this part of the problem as one in flat space $R^{1,d+1}$ with 
the understanding that the Ramond ground states in the cigar have charges and
 conformal weights different from those of two dimensional flat space in 
such a way that the ground states we write down are weight one states of the full string theory. 
The full set of zero mode equations we wish to solve is then:
\eqn\ferzmodes{
(\psi^{I}_{0} - i \eta \eta^{I}_{J} {\t \psi}^{J}_{0} )\ket{{\b Bp}}  =0, \quad I=0,1..d-1, d=\rho, d+1=\theta.
}

\ndt We choose the following representation of the gamma matrices for even
 dimension $d$ 
 \polchinski\ for $\IR^{1,d-1}$ 
\eqn\gammamatrices{\eqalign{
& \Gamma^{0} = \pmatrix{0 & 1 \cr -1 & 0} \qquad \Gamma^{1}  = \pmatrix{0 & 1 \cr 1 & 0} \qquad
\Gamma^{\mu} = \gamma^{\mu} \otimes \pmatrix{-1 & 0 \cr 0 & 1} \cr
& \Gamma^{d-2}  = I \otimes \pmatrix{0 & 1 \cr 1 & 0} \qquad \Gamma^{d-1}  =I \otimes \pmatrix{0 & -i \cr i & 0} \quad \qquad \mu=0,1,...,d-3. 
}}
We further define the parity and charge conjugation matrices:
$$\eqalign{
&\Gamma  = i^{1-{d\over 2}}\Gamma^{0}\Gamma^{1}...\Gamma^{d-1} \qquad
B_{1}  = \Gamma^{3}  \Gamma^{5}... \Gamma^{d-1} \qquad B_{2}= \Gamma B_{1} \cr
&C =B_{1} \Gamma^{0}\quad \hbox{if}\quad d=2\ \hbox{mod}\ 4\quad \hbox{and} \quad C  =B_{2} \Gamma^{0}\quad\hbox{if}\quad d=0\ \hbox{mod}\ 4 \,.
}$$
The operator $\Gamma$ is defined to have eigenvalues $\pm1$. The gamma matrices obey the
relations:
\eqn\gammacomm{\eqalign{
\left\{ \Gamma^{I}, \Gamma^{J} \right\} & = \eta^{IJ} = diag(-1,1,...1), \cr
\left\{ \Gamma, \Gamma^{I} \right\} & = 0, \quad (\Gamma^{I})^{t} = -C \Gamma^{I} C^{-1}. \cr
}}
Denoting the vacuum by $\ket{B}_{0}= \ket{A} \ket{\T B}$ where $A$ and $B$ are
the $2^{d+1 \over 2}$ dimensional spinor indices of $Spin(d+2)$, the action of
$\psi^{I}, \t \psi^{I}$ are those of the 
gamma matrices as implied by the fermion zeromode commutation relations:
\eqn\psiisgamma{\eqalign{
\psi^{I}_{0} \ket{A} \ket{\T B} & = \irt2 (\Gamma^{I})^{A}_{C} (1)^{B}_{D}  \ket{C} \ket{\T D} \cr 
\t \psi^{I}_{0} \ket{A} \ket{\T B} & = \irt2 (\Gamma)^{A}_{C}
(\Gamma^{I})^{B}_{D}  \ket{C} \ket{\T D}. \cr  
}}
Note that the gamma-matrix $\Gamma$ in 
the action of the right movers ensures that the left and right movers
anti-commute.
If we denote the solution of the zero-mode equation
\ferzmodes\ by $\CM_{AB} \ket{A} \ket{\T B}$, we can translate the equation
into one for the matrix of coefficients $\CM$:
\eqn\forM{
(\Gamma^{I})^{t} \CM - i \eta \, \eta^{I}_{J} \Gamma \CM \Gamma^{J} = 0.
}
A solution to the equation is: 
\eqn\solM{
\CM = C \Gamma {(1+ i \eta \Gamma) \over (1+i)}.
}

It is also useful to decompose the spinors into $d$ dimensional spinors with
 specific chirality under $\Gamma$. Since $\Gamma$ has eigenvalues $\pm1$, we
 can always choose a basis where the top half of the $2^{d \over 2}$ spinor
 and the bottom half have eigenvalues $\pm 1$. We can then decompose the
 matrix $\CM_{AB} = \pmatrix{M_{\a\b} & M_{\a\dot{\b}} \cr M_{\dot{\a}\b}
 &M_{\dot{\a}\dot{\b}} \cr}$. The vacuum solution can
 be written as a superposition: 
\eqn\vacsoln{
\ket{B, \eta = \pm}_{R} = \ket{\Omega}_{1} + i \eta \ket{\Omega}_{2} .
} 
where the two terms are given by:
\eqn\defomega{\eqalign{
 \ket{\Omega}_{1} & = M_{\a {\dot \b}} \ket{\a} \ket{\T{\dot \b}} \cr
 \ket{\Omega}_{2} & = M_{\dot \a \b} \ket{\dot \a} \ket{\t \b}. \cr 
}}
Here, we have chosen the $(-\half, -{3 \over 2})$ picture for the superghosts
 (which is a
useful to compute one point functions \yost \diVec). 
Notice that the two terms have definite fermion number eigenvalue: 
\eqn\etavacua{\eqalign{
(-)^{F}\ket{\Omega}_{1}  & = \ket{\Omega}_{1}, \quad (-)^{\T F}  \ket{\Omega}_{1} = \ket{\Omega}_{1}; \cr
(-)^{F}\ket{\Omega}_{2}  & = - \ket{\Omega}_{2}, \quad (-)^{\T F}  \ket{\Omega}_{2} = - \ket{\Omega}_{2}.
}}
which implies the relations:
\eqn\oddandevenvac{
(-)^{F}\ket{B, \eta}_{R} =  (-)^{\T F} \ket{B, \eta}_{R}  =  \ket{B, - \eta}_{R}. 
}
This completes the discussion of the solution to the zero-mode conditions on
 the boundary state in the R-sector.

\subsec{Spectral flowed extended characters}
The open string partition function for the point-like brane in the cigar
supercoset conformal field theory is most easily encoded in characters
of the $\CN=2$ superconformal algebra that are extended, i.e. summed over
spectral flow orbits. Below, the character 
$Ch_t$ denotes the extended character of the $\CN=2$ superconformal 
algebra associated to the trivial
representation of $SL(2,R)$. 
The trivial representation is a finite, one-dimensional
representation. We can associate a spin $j=0$ to this representation (or $u=1$
in the notation of \janbranes\ ). We recall
the trivial unextended characters:
\eqn\trivunext{\eqalign{
ch_t (r;\tau,\nu)\left[\matrix{0\cr 0}\right] &= q^{-{1\over 4k}+{r^2\over k}} z^{2r\over k}
{(1-q)\over (1+z q^{\half+r})(1+z^{-1} q^{\half-r})} 
{\theta_3(\tau,\nu)\over \eta(\tau)^3} 
\cr
ch_t (r;\tau,\nu)\left[\matrix{0\cr1/2}\right] &= q^{-{1\over 4k}+{r^2\over k}} (-z)^{2r\over k}
{(1-q)\over(1-z q^{\half+r})(1-z^{-1} q^{\half-r})} 
{\theta_4(\tau,\nu)\over\eta(\tau)^3} 
\cr
ch_t (r';\tau,\nu) \left[\matrix{1/2\cr 0}\right] &= q^{-{1\over 4k}+{r'^2\over k}} z^{2r'\over k}
{(1-q)\over (1+z q^{\half+r'})(1+z^{-1} q^{\half-r'})} 
{\theta_2(\tau,\nu)\over \eta(\tau)^3} 
}}
in the notation of  \janbranes\ (brought slightly closer to more standard 
notation). In the R-sector $r'=r+1/2$, i.e. it takes values in a range which is
shifted compared to the NS-sector range.
To define the extended characters, we perform a sum over spectral
flow orbits. 
Suppose we have a rational level $k = {N\over K}$, 
where $K,N$ are strictly positive integers (and let's suppose
they have greatest common divisor one).
We define the extended characters by a sum over spectral
flow orbits, determined by the integer $N$:
\eqn\extdefn{\eqalign{
Ch_t (r; \tau,\nu) \left[\matrix{a/2\cr b/2}\right] &=
\sum_{n \in N Z} q^{c n^2 \over 6 } z^{c n \over 3    } 
ch_t (r;\tau,\nu + n \tau) \left[\matrix{a/2\cr b/2}\right].
}}
We compute and find the following characters:
\eqn\extchar{\eqalign{
Ch_t (r; \tau,\nu) \left[\matrix{0\cr 0}\right] &=
\sum_{m \in Z} q^{-{K\over4N}} q^{KN(m+{r\over N})^2}
z^{2K(m+ {r\over N})}\cr 
& \left( {1\over 1+ z q^{mN+r+\half}}- {1\over 1+ z q^{mN+r-\half}} \right)
{\theta_3(\tau,\nu)\over \eta^3}  \cr
Ch_t (r; \tau,\nu) \left[\matrix{0\cr 1/2} \right]&= 
\sum_{m \in Z} q^{-{K\over 4N}} q^{KN(m+{r\over N})^2}
(-z)^{2K(m+ {r\over N})} 
\cr
&  \left( {1\over 1- z q^{mN+r+\half}}- {1\over 1- z q^{mN+r-\half}} \right)
{\theta_4(\tau,\nu)\over \eta^3} \cr
Ch_t (r'; \tau,\nu) \left[\matrix{1/2 \cr 0}\right] &=
\sum_{m \in Z} q^{-{K\over 4N}} q^{KN(m+{r'\over N})^2}
(z)^{2K(m+ {r'\over N})}
\cr
&  \left( {1\over1+ z q^{mN+r'+\half}}- {1\over1+ z q^{mN+r'-\half}} \right)
{\theta_2(\tau,\nu)\over \eta^3}.
}}
Again $r'=r+1/2$, i.e. $r'$ takes values in a shifted range compared to the
NS-sector where $r \in Z_N$.
For the twisted R-sector, generically, we need to be careful. Indeed,
although the $\theta_1$ function may be zero, there may appear a pole
in the denominator of the other factors, for particular values of $z$
and $r'$. We therefore note that generically we have:
\eqn\twistedr{\eqalign{
Ch_t (r'; \tau,\nu) \left[\matrix{1/2\cr1/2} \right] &= 
\sum_{m \in Z} q^{-{K\over 4N}} q^{KN(m+{r'\over N})^2}
(-z)^{2K(m+ {r'\over N})} \cr
& \qquad \left( {1\over 1- z q^{mN+r'+\half}}- {1\over1- z q^{mN+r'-\half}} \right)
{\theta_1(\tau,\nu)\over\eta^3}.
}}
The uses of these characters in the paper are as follows. First of all it is
important to realize in the light of the proof of the vanishing of the GSO
projected open
string partition function given in appendix
C
that the open string partition function can indeed be written as a trivial,
extended $N=2$ character, and in particular that it corresponds to a sum over
spectral flow orbits. This is manifest from the papers \EguchiIK \EguchiYI
\janbranes . Secondly, it is important to have the twisted R-sector partition
function at generic values of $z$, since it allows for the evaluation of
possible singular terms in the limit $z \rightarrow 1$ (as in the case of
$d=2$
in the bulk of the paper). Moreover, as observed in \EguchiIK \EguchiYI
\janbranes\ , the powerful formalism allows for easier generalization (to
orbifolds,
other levels, other backgrounds, etc). The above character formulas can
easily be
evaluated at the levels $k=2,1,2/3,1/2$ and yield (the coset
factors in) the open string
partition functions recorded in the bulk of the paper.

\appendix{B} {Closed String Vertex Operators on the Cigar}

In this section, we construct the primaries of the supersymmetric coset $SL(2,\IR)/U(1)$ at level $k$. We follow the conventions of \refs{\Giveon, \janbranes} and references therein. This theory has an $\CN=2$ superconformal symmetry. The parent supersymmetric $SL(2,\IR)_k$ theory has currents $J^a$ and $\psi^a$ which have coupled OPEs
\eqn\sltworcurrents{\eqalign{
J^a(z)J^b(0) &\sim {{k\over 2}g^{ab} \over z^2} + {f^{ab}_cJ^c}{z} \cr
J^a(z)\psi^b(0) &\sim {f^{ab}_{c}\psi^c \over z}
}}
with the fermions satisfying the usual OPEs. 
This theory is a product of a bosonic $SL(2,\IR)$ at level $k+2$, generated by the currents 
$$
j^a = J^a-\hat{J}^a = J^a - {i\over 2}f^{a}_{bc}\psi^b\psi^c \,,
$$
and three free fermions. The $U(1)$ symmetry to be gauged is generated  by $J^{3}, \psi_{cig}^{3}$.
The currents that make up the $\CN=2$ chiral algebra on the coset are:
\eqn\cigarcurrents{\eqalign{
T&=T_{SL_{2,\IR}} - T_{U(1)} \cr 
G^{\pm} &= (\psi_{cig}^{1} \pm i \psi_{cig}^{2}) (j^{1} \mp i j^{2}) \cr
J &= j^{3} - i\psi_{cig}^{1}\,\psi_{cig}^{2} \equiv i Q \p\theta + i \p H_{cig} \equiv i \p\phi \,.
}}

Now, the primaries of the bosonic $SL(2,\IR)$ (at level $k+2$) are denoted $V^{j}_{m_{bos}\mb_{bos}}$, where $m_{bos}$ is the charge under the purely bosonic $j^3$ current
$$
j^3(z)\Phi^{j}_{m_{bos}\mb_{bos}}(0) \sim {m_{bos}\Phi^{j}_{m_{bos}\mb_{bos}}\over z} \,.
$$
They have (left and right) conformal dimensions 
$$
\Delta(V_{j,m_{bos},\mb_{bos}}) = -{j(j+1)\over k} \,.
$$
As these fields are independent of the free fermions, they are also primary fields of the superconformal $SL(2,\IR)$ at level $k$. In order to obtain the primaries of the coset, it is useful to bosonize the various currents we have as follows:
\eqn\scalars{\eqalign{
\p H_{cig} &= i \psi_{cig}^{-}\psi_{cig}^{+}  \qquad  J^3 = -\sqrt{{k\over2}}\p X_3 \cr
J^R &= i\sqrt{{c\over3}}X_R  \qquad  j^3 = -\sqrt{{k+2\over2}}\p x_3 \,,
}}
where the normalizations ensure that the scalars have canonical OPEs. These scalars are not all independent and using the definition of the bosonic currents and the $\CN=2$ \cigarcurrents, we can rewrite all scalars in terms of $X_3$ and $X_R$:
$$
\sqrt{{k\over2}} x_3 = i X_R + \sqrt{{k+2\over 2}} X_3 \qquad \hbox{and} \qquad iH_{cig} = \sqrt{{2\over k}} X_3 + i \sqrt{{k+2\over k}} X_R   \,.
$$
Given these expressions, and knowing that the currents that are gauged in the coset are $J^3$ (in the bosonic case), we can decompose
\eqn\bosonicdecomp{
V^j_{m_{bos}} = \Phi^j_{m_{bos}} e^{m\sqrt{{2\over k+2}}x_3} \equiv \Phi^j_{m_{bos}} e^{2 m_{bos}\sqrt{{k+2\over k}}X_R}\, e^{m_{bos}\sqrt{{2\over k}}X_3} \,,
}
where $\Phi^j_{m_{bos}}$ is a primary of the bosonic Euclidean coset CFT (at level $k+2$). One infers that
$$
\Delta(\Phi^j_{m_{bos}}) = -{j(j+1)\over k} + {m_{bos}^2\over k+2}
$$
In the supersymmetric coset, we also gauge the fermionic current $\psi^3$ and the primary we start with in the parent theory is of the form $V^j_{m_{bos}} e^{inH}$. The coset primaries are found by using \bosonicdecomp\ and  
$$
e^{inH} = e^{n(\sqrt{{2\over k}}X_3+i\sqrt{{k+2\over k}}X_R)} \,.
$$
These two equations lead to the decomposition of the primary in the parent theory of the form
\eqn\fermionicdecomp{
V^j_{m_{bos}}\,e^{inH_{cig}} = \Phi^{j}_{m_{bos}}\, e^{i({2m_{bos}\over k+2}+n)\sqrt{{k+2\over k}}X_R}\, e^{\sqrt{{2\over k}}(m_{bos}+n)X_3} \,,
}
which allows one to infer that the superconformal coset primary is given by
$$
\Phi^{j,n}_{m_{bos}} = \Phi^j_{m_{bos}} e^{i({2m_{bos}\over k+2}+n)\sqrt{{k+2\over k}}X_R} \,.
$$
It is clear from equation \fermionicdecomp\ that the $J^3$ eigenvalue of the operator is given by 
$$
m = m_{bos}+n \,.
$$ 
In terms of $m$ and $n$, the conformal dimension is read off to be
$$
\Delta(\Phi^{j,n}_{m_{bos}}) = -{j(j+1)\over k} + {(m_{bos}+n)^2\over k} + {n^2\over 2} =  -{j(j+1)\over k} + {m^2\over k} + {n^2\over 2} \,,
$$
while the $R$-charge is given by 
$$
Q (\Phi^{j,n}_{m_{bos}})= {k+2\over k}\left({2m_{bos}\over k+2}+n\right) = {2m_{bos}\over k}+{n(k+2)\over k} = {2m\over k} + n\,.
$$
In the NSNS sector, we have $n\in \IZ$, while in the RR sector, we have $n\in \IZ+\half$. The full closed string field is obtained by putting together the left and right moving pieces and yields $\Phi^{j,n,\overline{n}}_{m_{bos},\mb_{bos}}$.
The axial gauging of the coset is done such that the $J^3$ and $\overline{J}^3$ eigenvalues $m$ and $\overline{m}$ are are related to the asymptotic momentum and winding of the circle direction of the cylinder at infinity\foot{The negative sign follows from the axial gauging of the coset.}
$$
m = {n+kw\over 2} \qquad \overline{m} = -{n-kw\over 2} \,.
$$ 
In the full theory, we tensor together these operators with the flat space operators as in \clstringops.
The physical operators are obtained by imposing BRST invariance and a GSO 
projection which tie together the two Hilbert spaces. 

\subsec{Reflection Amplitude}

For completeness, we also write down the reflection amplitude for the bulk fields. It is defined by the two point function of the fields
$$
\langle\Phi^{j,n,\nb}_{m_{bos},\mb_{bos}}(1) \Phi^{j',n'\nb'}_{m'_{bos},\mb'_{bos}}(0) \rangle = [\delta(j+j'+1)+R(j,m_{bos},\mb_{bos})\delta(j-j')] \, \delta_{m_{bos}+m_{bos}'}\,\delta_{\mb_{bos}+\mb_{bos}'}
$$
where
\eqn\reflectamp{
R(j,m_{bos},\mb_{bos}) = \nu^{2j+1} {\Gamma(2j+1)\Gamma(-j+m_{bos})\Gamma(-j-\mb_{bos})\Gamma(1+{2j+1\over k}) \over \Gamma(-2j-1)\Gamma(j+1+m_{bos})\Gamma(j+1-\mb_{bos})\Gamma(1-{2j+1\over k}) }\,.
}
Note that irrespective of whether the operators are in the NSNS or RR sector, the reflection amplitude is determined by the quantum numbers $m_{bos},\mb_{bos}$ from the $k+2$-level bosonic current algebra.

\appendix{C} {The construction of the conserved space-time supercharges}
In this appendix, we show that a combination of the supercurrents  $S_{\a}
(z)$ and $\t S_{\a}(\zb)$ is conserved on the open string worldsheet; in other
words, the supercharges $\int dz\,  \, S_{\a}(z) + \int d\zb \, \t S_{\a}(\zb)$
annihilate the boundary state. This construction is  subtle at the
quantum mechanical level \refs{\li , \calnap , \yost}. We shall  prove it at the classical level by studying the boundary conditions on the worldsheet fields.

As already discussed in Appendix B, the supersymmetric coset $SL_{2}/U(1)$ is described as a product of the bosonic $SL(2,\IR)_{k+2}$ and the two free fermions $\psi_{cig}^{\pm}$ which have OPE's which do not mix  \KazamaQP\ . Also, the fermions $\psi_{cig}^{a}$ have conformal weight $\half$ and can be bosonized as
\eqn\cigbosonize{
(\psi_{cig}^{1} + i \psi_{cig}^{2}) \equiv \Psi_{cig} = e^{i H_{cig}} \,.
}
We have already exhibited the $\CN=2$ algebra of the coset in \cigarcurrents. The $\CN=2$ supersymmetry in the $\IR^{1,3}$ factor can be seen by forming the linear combinations 
\eqn\cplxvar{\eqalign{
Z^{0} & = X^0 + X^1 \quad\quad \overline{Z}^0 = -X^0 + X^1 \quad\quad 
Z^{1} = X^{2} + i X^{3} \quad\quad  \overline{Z}^1 = X^2 -i  X^{3} \ldots  \cr
\Psi^{0} &= \psi^{0} + \psi^{1} \quad\quad \overline{\Psi}^{0} = - \psi^{0} + \psi^{1} \quad\quad
\Psi^{1} = \psi^{2} + i \psi^{3} \quad\quad \overline{\Psi}^{1} =  \psi^{2} -i  \psi^{3} \ldots 
}}
in terms of which
\eqn\mattercurrents{\eqalign{
T_{mat}  &= -{1\over 2} \p \overline{Z}^{a} \p Z_{a}  -{1\over 2}(\overline{\Psi}^{a}\p \Psi_{a} + \Psi^{a}\p \overline{\Psi}_{a})   \quad \quad
J_{mat}  = - \overline{\Psi}^a\Psi_a \cr
G^{+}_{mat} &=  i \Psi^{a}\p \overline{Z}_{a}  \quad \quad 
G^{-}_{mat} =  i \overline{\Psi}^{a}\p Z_{a} \quad\quad a= (0,\ldots, {d -2\over 2}) .
}}
We bosonize the fermions and denote them 
\eqn\bosonize{\eqalign{
\Psi^a & = e^{iH_a} \quad \overline{\Psi}^a = e^{- i H_a} \,.
}}
Denoting the bosonized superghost by $\varphi$, the dimension one operator which is the spacetime  supercharge is \kutsei
\eqn\supercharges{
S = e^{-{\varphi\over 2}}\,e^{{i\over 2}(H_1 + \dots + H_{cig} + Q \theta)} \,.
}
In order to verify that the boundary state we have constructed is supersymmetric (at least at the classical level), we will begin with the boundary conditions \ferzmodes\ 
\eqn\realferbc{
\left(\psi^{I}-i \eta\, \eta^I_J\, \t\psi^{J}\right)|_{\p\Sigma} = 0 ,
}
and construct the linear combination of \supercharges\ and its right-moving counterpart that is preserved by the D-brane. Rewriting this in the bosonized variables, we get the boundary conditions:
\eqn\Hbc{
\left(H_{a} = \t H_{a} + {\pi \eta \over 2}  \right) |_{\p\Sigma} , \quad  a= 0,\ldots, {d-2\over 2},cig . 
}
The superconformal ghosts obey the boundary condition  \calnap
\eqn\ghostbc{
\left( \varphi  =  \t{\varphi} + {i\pi \eta \over 2}  \right) |_{\p\Sigma} .
}
We need to supplement these standard boundary conditions with boundary conditions for the chiral boson $\theta$ which preserve the Neumann condition on the boson:
\eqn\thetabc{
\left( \theta = \t \theta + {\pi \eta Q \over 2 } \right) |_{\p\Sigma} .
}
We note as a check that for the case $d=6$, when the chiral boson is at the free fermion radius, the boundary condition \thetabc\ is the same as the one for the other fermions \Hbc . 

Using \Hbc , \thetabc\ and \ghostbc, one can check that the combination of supercharges that preserve the boundary condition is $\left(S+\t S\right)$. Other supercharges, which are local with respect to this one are obtained by flipping the signs in front of two of the three bosonized fields $H_a$. One can show that for all of these supercharges, the same left-right combination preserves the boundary condition.

\appendix{D}{A generic proof of the vanishing of supersymmetric open string
partition functions}
\seclab\gepnerproof
We sketch a generic proof of the vanishing of the GSO projected open string partition
function. Our aim in this section is to clarify conceptually the mechanism underlying the
delicate cancellation of the bosonic and fermionic contributions to the partition
functions. 
To lay bare the generic mechanism,
we export some important lessons from the construction of bulk supersymmetric
partition functions to our context (see especially  \gepner\  for a nice summary
of the relevant techniques in the bulk). The basic differences with the bulk
analysis
is that for the annulus amplitude, we needn't worry about modular invariance,
and on the other hand that we wish to prove the supersymmetry of the {\it open}
string spectrum -- we therefore apply the relevant part of the strategy of
\gepner\ in the open string channel directly.

Let's assume then, following \gepner\ that we are in light-cone gauge -- which
most straightforwardly encodes the physical spectrum. The (open string channel)
$N=2$ superconformal field theory has central charge $c=12$.
The (total transverse) $U(1)_R$ current can then be written as $J=i
2 \partial \phi$,
where $\phi$ is a canonically normalized scalar.
We define an associated operator (in the open
string channel):
$$
O_{1/2} = e^{i \phi} \,,
$$
which is the operator that implements spectral flow by half a unit.
The proof of the vanishing of the partition functions is based on two
assumptions.
Firstly, we assume that the partition function in the open string channel consists of
(supersymmetric)
characters which are of the following form:
$$
\chi_{susy} (\tau) = \sum_{n \in Z} (-1)^n O_{1/2}^n \chi_n(\tau)\,,
$$
namely, it is an alternating sum (over bosons and fermions) of contributions
that are related by half a unit of spectral flow, implemented by the 
operator $O_{1/2}$.

We observe that the degrees of freedom associated to the $U(1)_R$ scalar
 $\phi$
 may be fermionized using one (chiral) complex fermion. The complex
fermion will be in the R-sector, provided that all total light-cone $U(1)_R$
charges are odd. That is the second assumption. (The assumption is
realized by performing a GSO projection.)

We expect that the operators $O_{1/2}=e^{i \phi}$ and $O_{1/2}^{\dagger}=e^{- i
\phi}$, leave the 
total partition function invariant (up to a minus sign). For the first
operator,
this is manifest, while for the second it follows from the bijective
character of spectral flow (see also \gepner\ for a slightly different argument).

The vanishing of the partition function can now be argued for as follows.
We concentrate on the fermionized $U(1)_R$ scalar.
Decompose the $c=12$ theory into a $c=1$ theory of a complex fermion in the R-sector
and a theory with central charge $c=11$. The action of the operators $O_{1/2},O_{1/2}^{\dagger}$ is
on the first theory only, and it generates all states, starting from a
ground state (typically denoted $ \ket{s=- \half} $). 
The partition function in the open string sector will therefore be of the form:
$$
Z  =  \sum_i \chi_{susy}^i = \sum_i ({\theta_{1, 2}\over \eta} 
- {\theta_{-1,2}\over \eta}) Z^i_{c=11} = \sum_i  {\theta_{11}\over \eta} Z^i_{c=11} . 
$$
The first term in the third expression corresponds to  R-charges that are one modulo four,
while the second term corresponds to R-charges that are three modulo four.
The relative sign is necessary for the partition function to change sign under
the action of $O_{1/2}$.
The partition function is zero -- it has a factor that is zero, corresponding
to a twisted R-sector partition function for a chiral complex fermion.

In summary, the proof of the vanishing of the open string partition function
is generic, provided we can show that it is of the form we assumed above. It is
more straightforward then its closed string counterpart because we needn't
worry about modular invariance. The adaptation from the proof in \gepner\ is
 conceptual, in that we thought of it as applying to the open 
string channel.
Note that the above proof provides the rationale for
the theta-function identities proven in the bulk of the paper and  a lot of other identities
that may be
difficult to prove otherwise, and that it explains the generic logic for the
implementation
of the GSO projection in the open string channel.
 We note for instance that some identities for which numerical evidence was
provided in
\HikidaXU\  (referred to in \EguchiIK\ in a context very close to
that of our paper) can be proven in the above fashion. Note that the proof
equally well applies to the case of D-branes in compact Gepner models
 \RecknagelSB \GutperleHB\ . In all these cases, the conditions on the proof
are met.

\appendix{E}{Details of the partition functions of the various open string theories}

\subsec{d=4}

For the $D3$-brane in $d=4$, the sum over $s$ in \partnfnNS \ 
can be simplified as follows
$$\eqalign{
\sum_{s\in {1\over 2}+ \IZ} {1\over (1+q^s)} \left(q^{s^2-s} -q^{s^2+s}\right)
&= 
\left[\sum_{s \in -{1\over 2}-\IZ_+}\quad +\quad \sum_{s \in {1\over
2}+\IZ_+}\right] {1\over (1+q^s)} 
\left(q^{s^2-s} -q^{s^2+s}\right) \cr
& = \sum_{s \in {1\over 2}+ \IZ_{+}} (q^{s^2+s}-q^{s^2-s})\left({1\over (1+q^{-s})} - {1\over (1+q^s)}\right) \cr
&= \sum_{s \in {1\over 2}+ \IZ_{+}}\, q^{s^2-s}\, (1-q^s)^2 \cr
&= q^{-{1\over 4}}-2q^{{1\over 4}} + 2 q^{{3\over 4}} + \ldots  
}$$
Using the standard product representation for the $\Theta$-functions, we get 
$$\eqalign{
\left( {\Theta_{00}(it)\over \eta^3(it)}\right)^2 = q^{-{1\over
4}}\prod_{1}^{\infty} {(1+q^{m-{1\over 2}})^4\over {(1-q^m)^4}} 
= q^{-{1\over 4}}+4 q^{1\over 4}+10q^{3\over 4} + \ldots 
}$$
Multiplying the two contributions leads to\foot{We also plugged the modular
functions into a symbolic manipulation program and got the expansion to  
high order.} formula \NSfourdexp .

To prove the vanishing of the exact partition function, we use the 
identity (where the sum is a formal sum over any set), 
\eqn\sumid{\eqalign{
\sum_{s}  {q^{s^2-s} -q^{s^2+s}  \over 1\pm q^s} &= \sum_{s} q^{s^2-s}
\left({1 -q^{2s}  
\over 1 \pm q^s} \right) = \sum_{s} q^{s^2-s} (1 \mp  q^{s} ) \cr
&= \sum_{s} q^{s^2-s} \mp  \sum_{s}  q^{s^{2}} = q^{-{1\over 4}}\sum_{s}
q^{(s-\half)^2} \mp  
\sum_{s}  q^{s^{2}} 
}}
We can then re-express 
\eqn\reexpress{\eqalign{
A^{NS} & =  \left( {\Theta_{00}(\tau)\over \eta^3(\tau)} \right)^{2} 
\left( - \Theta_{10}(2\tau) + e^{-{i \pi \tau \over 2}}  \Theta_{00}(2\tau) \right) \cr
A^{\T NS} & = \left( {\Theta_{01}(\tau)\over \eta^3(\tau)} \right)^{2} 
\left( \Theta_{10}(2\tau) + e^{-{i \pi \tau \over 2}}  \Theta_{00}(2\tau) \right) \cr
A^{R} & = \left( {\Theta_{10}(\tau)\over \eta^3(\tau)} \right)^{2} 
\left( - \Theta_{00}(2\tau) + e^{-{i \pi \tau \over 2}}  \Theta_{10}(2\tau) \right) \cr
A^{\T R} & = 0.
}}

\subsec{d=6}

We use a formal power series expansion to re-express the first term $Z^{NS}$ 
as\foot{It is dangerous to use the formal power series expansion for
$s<0$. In a  more careful analysis, one splits the sum over $s$ into two sums, both containing positive powers of $q$ as we did previously.}:
\eqn\reexpsixd{\eqalign{
Z^{NS} & =  {\Theta_{00}^{3}(\tau)\over \eta^9(\tau)}  \sum_{s \in \IZ + \half} \sum_{r=0}^{\infty} (-1)^{r}  q^{rs} q^{-{1 \over 8}} \left(q^{\half(s-\half)^{2}} - q^{\half(s+\half)^{2}} \right)  \cr 
& = {\Theta_{00}^{3}(\tau)\over \eta^9(\tau)}  \sum_{r=0}^{\infty} (-1)^{r}    q^{{r \over 2}-{1 \over 8}} 
\sum_{s \in \IZ + \half} q^{r (s-\half)} \left(    q^{\half(s-\half)^{2}} -   q^{\half(s+\half)^{2}} \right) \cr
& = {\Theta_{00}^{3}(\tau)\over \eta^9(\tau)}  \sum_{r=0}^{\infty} (-1)^{r}    q^{{r \over 2}-{1 \over 8}} 
\sum_{n \in \IZ} \left(   q^{\half n^{2}+nr} -    q^{\half (n+1)^{2} + nr} \right) \cr
& = {\Theta_{00}^{3}(\tau)\over \eta^9(\tau)}  \sum_{r=0}^{\infty} (-1)^{r}    q^{{r \over 2}-{1 \over 8}} 
\left(\sum_{m \in \IZ} q^{\half m^{2}}\right) \left(q^{-{r^{2} \over 2}}-    q^{-{r^{2} \over 2}-r} \right) \cr
& = {\Theta_{00}^{4}(\tau)\over \eta^9(\tau)}  \sum_{r=0}^{\infty} (-1)^{r}  \left(q^{-\half (r-\half)^{2}}-  q^{-\half (r+\half)^{2}}   \right) \cr
}}
Using similar manipulations, we get the other the equations in \sixdpfnparts .

\listrefs
\end